\documentclass[a4paper,11pt]{article}
\pdfoutput=1 

\usepackage{jheppub} 
\usepackage{amsmath,mathrsfs}
\usepackage[T1]{fontenc} 
\usepackage{multirow}
\usepackage[dvipsnames]{xcolor}
\def\be{\begin{equation}}
\def\ee{\end{equation}}
\def\bea{\begin{eqnarray}}
\def\eea{\end{eqnarray}}

\def\yzero{\smash{\hbox{$y\kern-4pt\raise1pt\hbox{${}^\circ$}$}}}

\def\beq{\begin{equation}}
\def\eeq{\end{equation}}
\def\beqa{\begin{eqnarray}}
\def\eeqa{\end{eqnarray}}

\def\-{\hphantom{-}}

\def\s2{\frac{1}{\sqrt2}}

\def\beq{\begin{equation}}
\def\eeq{\end{equation}}
\def\beqa{\begin{eqnarray}}
\def\eeqa{\end{eqnarray}}

\def\IF{\relax{\rm I\kern-.18em F}}
\def\II{\relax{\rm I\kern-.18em I}}
\def\IP{\relax{\rm I\kern-.18em P}}
\def\IC{\relax\hbox{\kern.25em$\inbar\kern-.3em{\rm C}$}}
\def\IR{\relax{\rm I\kern-.18em R}}

\def\Dsl{\,\raise.15ex\hbox{/}\mkern-13.5mu D} 
\def\IZ{Z\kern-.4em  Z}

\def\sig{\sigma}

\title{\boldmath Spinning solutions for the bosonic M2-brane with $C_{\pm}$  fluxes}

\author[a]{P. D. Alvarez,}
\author[b]{P. Garcia,}
\author[a]{M.P. Garcia del Moral,}
\author[a,c]{J.M. Peña,}
\author[a]{and R. Prado\footnote{The order of the authors is alphabetical.}}

\affiliation[a]{Departamento de F\'isica, Universidad de Antofagasta,  Avda. Universidad de Antofagasta 02800, Antofagasta, Chile.}
\affiliation[b]{Laboratorio de Sistemas Complejos, Facultad de Ingeniería, Universidad Central de Venezuela, Los Chaguaramos 
cp. 1050, Caracas, Venezuela.}
\affiliation[c]{Departamento de F\'isica, Universidad Católica del Norte, Avda
Angamos 0610, Antofagasta, Chile}
\emailAdd{pedro.alvarez@uantof.cl}
\emailAdd{pedro@fisica.ciens.ucv.ve}
\emailAdd{maria.garciadelmoral@uantof.cl}
\emailAdd{joselen@yahoo.com}
\emailAdd{reginaldo.prado@ua.cl}

\abstract{ In this work we obtain classical  solutions of the bosonic sector of the supermembrane theory with two-form fluxes  associated to a quantized constant $C_{\pm}$ background. This theory satisfies a flux condition on the worldvolume that induces monopoles over it. Classically it is stable as it does not contain string-like spikes with zero energy in distinction with the general case. At quantum level the bosonic membrane has a purely discrete spectrum but the relevance is that the same property holds for  its supersymmetric spectrum. We find for this theory spinning  membrane solutions, some of them including the presence of a non-vanishing symplectic gauge connection defined on its worldvolume in different approximations. By using the duality found between this theory and the so-called supermembrane with central charges,  rotating membrane solutions found in that case, are also solutions of the M2-brane with $C_{\pm}$ fluxes. We generalize this result to other embeddings. We  find new distinctive rotating  membrane solutions, some of them including the presence of a non-vanishing symplectic gauge connection defined on its worldvolume. We obtain numerical and analytical solutions in different approximations characterizing the dynamics of the membrane with fluxes $C_{\pm}$ for different ansätze of the dynamical degrees of freedom. Finally we discuss the physical admissibility of some of these ansätze to model the components of the symplectic gauge field.}

\begin{document}

\maketitle
\flushbottom

\newpage
\section{Introduction}
Classical solutions of membrane theory can be useful to understand better the M-theory properties of its fundamental elements. Other historical interest in this study has been focused in the context of a generalization to M-theory of the nonrelativistic strings formulation on AdS spaces, see for example \cite{Hoppe2004SpinnigMembraneAdS}.

Rotating membrane solutions are compelling studies for a variety of reasons:  They can  be sources of supergravity solutions in eleven dimensions that can describe charged rotating black holes in lower dimensions, see for example \cite{CveticChong2005,Klemm2011}, they can provide a preliminary signal for interpreting membranes as extended spinning particle models for those cases when they become well-defined at supersymmetric quantum level on a proper background, or they can even describe spinning solitonic solutions. The existence of membrane solitonic solutions was discussed in several works. For example in terms of solitary waves in \cite{Restuccia:1998_MembraneSolitons},  in the context of Q-ball matrix model \cite{Soo-Jong_Rey}, as instantonic solutions \cite{FLORATOS:1998_InstantonSolutionsSelf-dualMembrane}, or in terms of membranes formulated on hyperkähler backgrounds \cite{Bergshoeff:1999_SolitonsOnSupermembrane,Portugues_2004}.  

In the literature the study of the classical solutions of the membrane theory firstly appeared in \cite{nicolaihoppe}. It was shown that spherical and toroidal membrane solutions could be  obtained from the membrane when it is embedded on spherical backgrounds. In the context of matrix models it was discussed in \cite{Hoppe:1997SolutionsMatrixModelEquations}. Spinning solutions for the membrane were proposed in \cite{FloratosRotatingToroidalB2002ranes} formulated in the Light Cone Gauge (LCG). Other spinning solutions for the 11D membrane matrix model on a pp-wave, i.e. in the context of BMN matrix model, were found in \cite{ArnlindHoppe2003,AXENIDESFloratos2017}.  
Rotating membrane solutions with the conserved charges in M-theory were studied in a series of papers \cite{Bozhilov:2005MembraneSolutions} in  the context of G2 or AdS manifolds \cite{Bozhilov2006ExactRotatingMembraneSolutionsG2manifold,Hartnoll_2003,ARNLIND2004118SpinningMembranes, BOZHILOV2008429} also in the presence of global $U(1)$ symmetry \cite{TRZETRZELEWSKI2009523}. The formulation of spinning membranes toroidally compactified on $M_9\times T^2$ appeared in \cite{FloratosRotatingToroidalB2002ranes,BRR2005nonperturbative}. 
This last one is the set-up that we will analyze in this paper for a background with fluxes $C_{\pm}$.

Recently it has been observed that the supermembrane theory toroidally compactified $M_9\times T^2$ on a background with a particular choice of quantized constant three-form components $C_{\pm}$ \cite{mpdm:2019:flux}. It induces a two-form flux on the target torus whose pullback generates a two-form flux over the worldvolume of the supermembrane. This condition implies the existence of a central charge condition on the worldvolume. Furthermore this theory is equivalent modulo a constant shift, or 'dual' to a toroidally compactified supermembrane  with central charges. The supermembrane with central charges was studied in \cite{Ovalle1,Ovalle2} and it is responsible for the generation of new terms in the Hamiltonian that render the supersymmetric spectrum purely  discrete with finite multiplicity from $[0,+\infty)$ as rigorously proved in \cite{BOULTON-2003-Discreteness}. Consequently, it can represent part of the microscopical degrees of freedom of the M-theory. It is a well-known fact that the supermembrane on $M_{11}$ has a continuous spectrum \cite{dwln} and this behaviour does not change simply by compactifying the manifold. 

The discreteness property of the supermembrane in the presence of fluxes $C_{\pm}$ that we refer is -up to now-  a condition particular to the cases considered. The spectra of both theories (the one with a central charge condition and the one with fluxes $C_{\pm}$) only  differ on a constant shift. We believe that the understanding of the dynamics of the bosonic sector of these well-behaved supermembranes,  is an important step towards its full-fledged characterization. A characteristic of these membranes, different from the usual ones is that  classically they do not contain string-like spikes with zero cost energy \cite{MPGMZ:Restuccia:2002SpectrumNoncommutativeD11}, hence classically they admit an interpretation as extended objects with preserved topology that do not split into pieces.

In this work we obtain some solutions to the equations of motion (E.O.M) associated to the bosonic sector of the supermembrane (also denoted along the paper as membrane or M2-brane)  toroidally compactified on $M_9\times T^2$ formulated on the LCG on a $C_{\pm}$ background. This simple target space still captures all of these new features. The membrane with fluxes has a symplectic gauge field and a nontrivial $U(1)$ gauge over its worldvolume.

 We compare our results for a vanishing symplectic gauge connection with those of \cite{BRR2005nonperturbative} adding the topological flux condition. This condition is equivalent to impose the so-called central charge restriction { \cite{MARTINRestucciaTorrealba1998:StabilityM2Compactified}}. The central charge condition is a geometrical condition imposed on the wrapping of the membrane around the compactified target-space that induces the presence of monopoles over the worldvolume and generates a central charge in the supersymmetric algebra. 
 
 The dynamical equations for the M2-brane with fluxes are a system of nine non-linear coupled partial equations highly nontrivial plus three constraints.  Moreover, it is known that those equations may admit soliton solutions when they are formulated on certain backgrounds. 
 In our analysis we also consider  other solutions, that we denote as 'Q-ball like', which admit $U(1)$ Noether charges  and for the approximations considered they exhibit a discrete spectrum. These results open  a window for a future  study -outside of the scope of this paper- to characterize further these solutions in order to determine whether they can truly model Q-ball solitons.
 
 The paper is structured as follows: In section \ref{s2}. we review the formulation of the supermembrane toroidally wrapped on $M_9\times T^2$ subject to a $C_{\pm}$ flux condition. We explain its relation with the formulation of the supermembrane compactified on the same target space with a central charge condition associated to an irreducible wrapping.  From section \ref{s3}. to section \ref{S7}. we present our results. In section \ref{s3}. we obtain the E.O.M of the (bosonic) membrane on a background $C_{\pm}$ and we discuss the type of ansätze that we will explore. In section \ref{s4}. we obtain explicit spinning solutions for the M2-brane with fluxes when formulated in a particular embedding in the absence of of a symplectic gauge field. We also obtained the mass operator of the M2-brane with fluxes $C_{\pm}$ formulated in this background  and we show that it contains the energy operator found in \cite{BRR2005nonperturbative}. In section \ref{s5}.  we characterize new solutions for the toroidal membrane that we denote as 'Q-ball like', for zero symplectic gauge field. We perform a numerical analysis and we obtain the first eigenvalues and eigenfunctions of its discrete spectrum for different boundary conditions. We discuss the discrete landscape and structure of the solutions with and without central charge.
 In section \ref{s6}. we obtain new solutions to the approximate equations that also include a non-vanishing gauge field. We discuss several cases: a first one, more restrictive, in which the approximation is imposed on both of the dynamical degrees of freedom, i.e. the complex scalar fields and the functions associated to the components of the gauge symplectic field, a second case in which the approximation is only imposed on one of the complex scalar fields, allowing to have a mixed dynamical system, and a third one in which the approximation is only imposed on the functions associated to the functions associated to the symplectic gauge field. We discuss the solutions on each case. In section \ref{S7}. we consider the full-fledged system of nonlinear PDE equations. We obtain mathematical exact solutions for the case with $Z_a$ with $a=1,2,3$ constant, and $Z_a$ a spinning ansatz. In section \ref{7.3} we discuss why these solutions in spite of solving exactly a complex nonlinear seven coupled equation system they are non-physical since their associated one-form does not fulfill the physical requirement of exactness required to describe the symplectic gauge field. We provide approximate admissible physical solutions allowing a symplectic gauge field  for the  $Z_a$ ansätze considered. In section \ref{s8} we present our discussion and conclusions.

\section{The toroidally compactified M2-brane with \texorpdfstring{$C_{\pm}$}{} fluxes}
\label{s2}
We review recent results obtained in \cite{mpdm:2019:flux} and \cite{mpgm10Monodromy} in the  supermembrane theory in the L.C.G. formulated formulated on a $M_9\times T^2$ background  with two-form fluxes induced by the presence of a quantized constant bosonic 3-form gauge fields $C_{\mu \nu \lambda }$. 
 
 This is a consistent background of the  supergravity equations of motion. The M2-brane is considered as a probe. When restricted to this background, the action of a supermembrane  becomes greatly simplified  \cite{mpdm:2019:flux}
{\small
\begin{equation}
\begin{aligned}
\label{2.6}
S= & - T \int d^3 \xi \{ \sqrt{-g}+\varepsilon^{uvw}\bar{\theta}\Gamma_{\mu \nu}\partial_w \theta \left[ \frac{1}{2}\partial_u X^\mu (\partial_v X^\nu\right. +  \bar{\theta}\Gamma^\nu \partial_v\theta) + \\
& + \frac{1}{6}\bar{\theta}\Gamma^\mu \partial_u \theta \bar{\theta}\Gamma^\nu \partial_v \theta ] +\frac{1}{6}\varepsilon^{uvw}\partial_u X^\mu \partial_v X^{\nu} \partial_w X^\rho C_{\rho \nu \mu} \} \, .
\end{aligned}
\end{equation}
}
where the $(X^{\mu}(\xi)$ represents the embedding coordinates and $\xi^u $ the worldvolume coordinates. $\theta^{\alpha}(\xi)$ represents a Majorana spinor of 32 components that transforms as an scalar under reparametrizations in the worldvolume. $\mu, \nu, \lambda $ denote the bosonic target space indices awhile $u,v,w$ denote the worldvolume ones. 

This supergravity background is an asymptotic limit of a  supergravity background originally discovered by  \cite{DUFF1Stelle991113,KStelleLectures} generated by a M2-brane acting as a source. The metric is given by 
{\small
\begin{equation}
\label{metric-stelle}
ds^2=(1+\frac{k}{r^6})^{-\frac{2}{3}}{dx^{\bar{\mu}}}{dx^{\bar{{\nu}}}}{\eta_{{\bar{\mu}} {\bar{\nu}}}} + (1+\frac{k}{r^6})^{-\frac{1}{3}}{dy^{\bar{m}}}{dy^{\bar{{n}}}}{\delta_{{\bar{m}} {\bar{n}}}} \, ,
\end{equation}}
for ${\bar{\mu}}= 0, 1, 2$, ${\bar{{m}}} = 3, ... , 10$ and $r = \sqrt{y^{\bar{{m}}}y^{\bar{{m}}}}$ the radial isotropic coordinate in the transverse space. The 3-form in this background takes the form,
{\small
\begin{equation}
\label{3form-stelle}
C_{{\bar{\mu}}{\bar{\nu}}{\bar{\sigma}}} = \epsilon_{{\bar{\mu}}{\bar{\nu}}{\bar{\sigma}}} (1+\frac{k}{r^6})^{-1} \, ,
\end{equation}}
 When $r\rightarrow \infty $, the metric (\ref{metric-stelle}) goes to Minkowski metric and (\ref{3form-stelle}) is constant. This is the background that we will consider from now on. 

We now consider the supermembrane action in the Light Cone Gauge (LCG) on a $M_{11}$ target space with constant gauge field $C_{{{\mu}}{{\nu}}{{\lambda}}} $ closely following the definitions in \cite{deWit}. The supersymmetric action is \cite{mpdm:2019:flux},
{\small
\begin{equation}
S = T\int d^3 \xi \{ -  \sqrt{\bar{g}\Delta}-\varepsilon^{uv}\partial_uX^a \bar{\theta} \Gamma^- \Gamma_a \partial_v \theta +  C_+ +\partial_{\tau} X^- C_- +\partial_{\tau} X^a C_a+C_{+-} \}
\end{equation}
}
with\footnote{In order to be self-contained we include the definitions 
of  \cite{dwhn}:
$\Delta = -g_{00} + u_r {\bar{g}}^{rs} u_s$ being ${\bar{g}}^{rs} g_{st} = {{\delta}^r}_t$ and $g \equiv det g = - \Delta \bar{g}$  (with $\varepsilon^{0rs}=\varepsilon^{rs}$).
$
{\bar{g}}_{rs} \equiv g_{rs} = \partial_r X^a \partial_s X^b {\delta}_{ab} ;\quad
$
$u_r \equiv g_{0r} = \partial_r X^- + \partial_0 X^a \partial_r X^b {\delta}_{ab} + \bar{\theta}{\Gamma}^{-} \partial_r \theta ;$
${\bar{g}}_{ 00} = 2\partial_0 X^- + \partial_0 X^a \partial_0 X^b {\delta}_{ab} + 2\bar{\theta}{\Gamma}^{-} \partial_0 \theta.
$}
{\small
\begin{equation}
\begin{aligned}
& C_a  =  -\varepsilon^{uv}\partial_uX^- \partial_vX^b C_{-ab} +\frac{1}{2}\varepsilon^{uv}\partial_uX^b \partial_vX^c C_{abc} \, , \\
& C_{\pm}  =  \frac{1}{2}\varepsilon^{uv}\partial_uX^a \partial_vX^b C_{\pm ab} \,, \qquad C_{+-}  =  \varepsilon^{uv}\partial_uX^- \partial_vX^a C_{+-a} \,,
\end{aligned}
\end{equation}
}
where
$a,b,c=1,...,9$ label the target space transverse coordinates indices, and $u,v=1,2$ label the space-like  worldvolume coordinates $(\sigma^1,\sigma^2)$. It is possible to fix the variation of some components of the 3-form by gauge invariance. In particular it is possible to fix $C_{+-a}=0$ and $C_{- ab}=0.$  We choose a background with $C_{- ab}=constant$ different from zero.

By performing a Legendre transformation the following Hamiltonian is obtained
{\small
\begin{equation}
H=T\int{d^2 \sigma}\{\frac{1}{(P_ -- C_-)}\left[\frac{1}{2}(P_a-C_a)^2  +\frac{1}{4}\left(\varepsilon^{rs}\partial_r X^a\partial_s X^b\right)^2\right] + \varepsilon^{rs}\bar{\theta}\Gamma^- \Gamma_a \partial_s \theta \partial_r X^a -C_{+}\}\,
\end{equation}}
subject to the primary constraints
%
{\small\begin{equation}
\label{ecuacion10constraints}
P_a \partial_r X^a + P_-\partial_r X^- + \bar{S}\partial_r \theta \approx  0 \quad ,  \quad
S+(P_- -C_-)\Gamma^{-}\theta \approx  0\,.
\end{equation}
}
 In order to define the physical Hamiltonian $X^-$ must be eliminated. In \cite{mpdm:2019:flux} the dependence on $X^-$ was eliminated by performing a canonical transformation on the configuration variables without introducing non-localities. On the new variables the Hamiltonian of the compactified theory on $M_9 \times T^2$ target space is the following one:
 {\small
\begin{equation}
\label{hamiltonian-fluxes}
\begin{aligned} 
\widetilde{H} = T\int_{\Sigma} d^2\sigma \{ \frac{\sqrt{w}}{\hat{P}_{-}^0} \, \left[ \frac{1}{2} \left(\frac{P_m}{\sqrt{w}} \right)^2 + \frac{1}{2}\left(\frac{P_i}{\sqrt{w}} \right)^2 + \frac{1}{4} \left\{ X^i, X^j \right\} ^2 +  \frac{1}{2} \left\{ X^i, X^m \right\}^2  \right. \\
\left. + \frac{1}{4} \left\{ X^m, X^n \right\} ^2 \right] +\sqrt{w} \left[ \bar{\theta}\Gamma^-\Gamma_m\left\lbrace X^m,\theta\right\rbrace +\bar{\theta}\Gamma^-\Gamma_i\left\lbrace X^i,\theta\right\rbrace\right]  -C_+ \} \,,
\end{aligned}
\end{equation}
}
where the  $X^m, m=3,\dots,9$ denote the transverse maps from the foliated worldvolume $\Sigma$ to $M_9$ and $X^i, i,j=1,2$ the maps  from $\Sigma$ to $T^2$ and the Lie bracket is defined as  $\{A,B\}=\frac{\epsilon_{uv}}{\sqrt{w}}\partial_uA\partial_v B$. In the compactified case, in contrast to the noncompact one, the last term in (\ref{hamiltonian-fluxes}) for constant bosonic 3-form is a total derivative of a multivalued function, therefore its integral is not necessarily zero.
This Hamiltonian (\ref{hamiltonian-fluxes}) is subject to the local and global constraints associated to the area preserving diffeomorphisms (APD) connected to the identity 
{\small
\begin{equation}
\label{constraintdos}
d(P_i dX^i + P_m dX^m+\overline{\theta}\Gamma^{-}d\theta)=0 \,, \qquad
 \oint_{{\mathcal{C}}_s} (P_i dX^i + P_m dX^m+\overline{\theta}\Gamma^{-}d\theta)=0 \,.
\end{equation}
}
In \cite{mpdm:2019:flux} it was found the Hamiltonian formulation of a supermembrane in the LCG compactified on $M_9\times T^2$ with a nontrivial two-form flux background induced by the presence of constant quantized components of the target space three-form,
 $C_{\pm}$, where
\begin{equation}
C_{\pm} = \frac{1}{2}\varepsilon^{uv}\partial_uX^{r} \partial_vX^{s} C_{\pm rs} \ , 
\end{equation}
and $C_{\pm rs}=\epsilon_{rs}c_{\pm}$ with $c_{\pm}\ne 0$ is a constant coefficient, such that it has an  associated 2-form flux background $F_2^{Back}$ 
\begin{equation}
    \int_{T^2}C_{\pm}=\int_{T^2} F_2^{Back}=k\in \mathbb{Z}, \quad k\ne 0.
\end{equation}
 Upon toroidal compactification  the background still corresponds to the asymptotic limit of a supergravity solution. The quantization condition on the three-form implies the existence of a topological condition on the background associated to the presence of a two-form flux condition over the torus whose pullback on the worldvolume acts as an extra constraint on the Hamiltonian. Due to the flux condition, it is no longer possible to perform changes of the three-form that could violate it, providing stability to the classical solutions. 
 
 In \cite{Ovalle1, Ovalle2}
the supermembrane with central charges associated with an irreducible wrapping was obtained. It corresponds to a M2-brane formulated on  $M_9\times T^2$ subject to a topological condition associated to an irreducible wrapping. It has the distinctive property of having a purely discrete spectrum with eigenvalues of finite multiplicity as rigorously shown in \cite{BOULTON-2003-Discreteness}. In the following we will shorten it by MIM2  since it represents a supermembrane minimally immersed in the background. 
The embedding maps $X^m(\tau,\sigma,\rho)$ on the non-compact space are defined as  $X^m(\tau,\sigma,\rho) :\Sigma \to M_9$ with $m=2,...,8$ and respectively the $X^{r}(\tau,\sigma,\rho) :\Sigma \to T^2$, where $r=9,10$,  describe the embedding on the compactified 2-torus. The coordinates of the supermembrane worldvolume are denoted by $(\tau,\sigma,\rho)$ parametrizing $\Sigma \times \mathbb{R}$ with $\Sigma$ denoting a Riemann surface of genus one and $\mathbb{R}$ parametrizing  the time. The maps $X^{r}$ satisfy the standard winding condition
\begin{equation}
    \oint_{\mathcal{C}_s}dX^{r}=R^{r}m_{r}^{s} \ ,
\end{equation}
where $m_{r}^{s}$ are the winding numbers and $R^{r}$ the torus radii. The MIM2 is subject to an irreducible wrapping condition 
\begin{equation}
\label{cargaCentral}
\int_{\Sigma} d X^{r} \wedge d X^{s}=\epsilon^{r s} n A_{T^2}, \quad n \in \mathbb{Z} /\{0\}\ , 
\end{equation}
where $A_{T^2}$ represents the 2-torus target space area. The above condition is responsible for the appearance of a non-vanishing central charge in the supersymmetric algebra.
This condition implies that the one-forms associated with the embedding map of the compact sector, can be globally decomposed by a Hodge decomposition  as follows $d X_{r}(\sigma,\rho,\tau) = d{X}_{rh}(\sigma,\rho) + dA_{r}(\sigma,\rho,\tau)$, with $d{X}_{rh}=R^{r}m_{r}^{s} d\widehat{X}_s(\sigma,\rho)$ a closed one-form defined in terms of the harmonic forms $d\widehat{X}_s$ and $dA_{r}$ an exact one-form. In {\cite{MARTINRestucciaTorrealba1998:StabilityM2Compactified}} It was  shown that  the integer $n$ associated with the central charge condition is  $n=det\mathbb{W}$ where $\mathbb{W}$ is the winding matrix. The central charge condition corresponds to a monopole condition given by the curvature $\widehat{F}=\epsilon_{rs}d X^{r}_h \wedge d X^{s}_h$ associated to a nontrivial $U(1)$ gauge field $\widehat{F}=d\widehat{A}$  defined on the membrane worldvolume. The bosonic LCG Hamiltonian of the theory corresponds to {\cite{Ovalle1,Ovalle2}}
\begin{equation}
\begin{aligned}
\label{hamiltonianirred}
H_{MIM2}&=T^{-2/3}\int_\Sigma d^2\sigma \sqrt{W}\Big[\frac{1}{2}\Big(\frac{P_m}{\sqrt{W}}\Big)^2+\frac{1}{2}\Big(\frac{P_{r}}{\sqrt{W}}\Big)^2\Big]\\ 
&+ T^{-2/3}\int_\Sigma d^2\sigma \sqrt{W}\Big[ \frac{T^{2}}{4}\left\{X^m,X^n\right\}^2 + \frac{T^{2}}{2}(\mathcal{D}_{r} X^m)^2+\frac{T^{2}}{4}(\mathcal{F}_{rs})^2 \Big],
\end{aligned}
\end{equation}
where $T$ is the tension of the the membrane and $W$ is the determinant of the induced spatial part of the foliated metric on the membrane, $\left\{ A,B\right\}=\frac{1}{\sqrt{W}}\epsilon^{ab}\partial_a A\partial_b B;\ $ is the symplectic bracket with $a,b=1,2$ and $\partial_a=\partial_{\sigma},\partial_{\rho}$ and  $\sqrt{W}=\frac{1}{2}\epsilon^{ab}\epsilon_{rs}\partial_a\widehat{X}\partial_b\widehat{X}$ defined in terms of the harmonic one-forms $d\widehat{X}$ of the Riemann surface.
The canonical momentum associated to the scalar fields $X^m,A_{r}$ respectively are $P_m$ y $P_{r}$. Due to the imposition of the central charge condition, there exists a new  dynamical degree of freedom  $A_{r}$ whose one-form $\mathbb{A}=dA$ transforms as a symplectic connection under symplectomorphisms. This gauge field is not present on a toroidal M2-brane formulation without this condition. The symplectic derivative is defined as
  $\mathcal{D}_{r}\bullet= D_{r}\bullet+\{A_{r},\bullet\}$  with $D_{r}\bullet=2\pi m_{r}^u\theta_{uv} R _{r}\frac{\epsilon^{{a}{b}}}{\sqrt{W}}\partial_{{a}}\widehat{X}^v\partial_{{b}}\bullet.$ \,
 The derivative $D_{r}$ is 
 defined in terms of the moduli of the  2-torus,  the harmonic one-forms $d\widehat{X}_{r}$, and one matrix $\theta_{uv}$ with $u,v=9,10$, related to the monodromy associated to its global description in terms of a torus bundle \cite{GMPR2012}. 
Therefore, the Hamiltonian contains new terms associated to the  symplectic covariant derivative of the scalar embedding maps $\mathcal{D}_{r} X^m$ and a symplectic curvature $\mathcal{F}$ defined by
\begin{equation}
 \mathcal{F}_{rs}= D_{r} A_s -D_s A_{r} +\left\{ A_{r} ,A_s \right\}.  \,
\end{equation}
The constraints of the theory associated to the local Area Preserving Diffeomorphism (APD) are:
\begin{equation}\label{APD local}
\mathcal{D}_{r} P_r+\left\{X^m,P_m\right\}\approx 0\ ,
\end{equation}

and to the APD global constraints, 
\begin{equation}\label{APD global}
\oint_{\mathcal{C}_s} \Big(\Big(\frac{P_{r}}{\sqrt{W}}\Big)\partial_a X^{r}+\Big(\frac{P_m}{\sqrt{W}}\Big)\partial_a X^m\Big)d\sigma^a \approx 0\ .
\end{equation}
In \cite{mpdm:2019:flux} it was proved that  supermembrane with $C_{\pm}$ fluxes  is in one-to one correspondence with the so-called supermembrane with a central charge condition associated to an irreducible wrapping modulo a constant shift. Indeed, it was shown that  
\begin{equation}
   \int_{T^2} F_2^{Back}=\int_{\Sigma} \widehat{F}.
\end{equation}
Hence in the following we will refer indistinctly to this condition as the flux quantization condition or the central charge condition\footnote{ Strictly speaking the induced $C_{\pm}$ worldvolume flux is proportional to the central charge condition as $c_{\pm}n$, for $c_{\pm}=1$ the equality holds. One can always redefine the wrapping numbers to absorb this factor without altering the results.}.
The Hamiltonian formulation of a supermembrane on this nontrivial quantized $C_{\pm}$ background  corresponds to 
\begin{equation}
    H_{C_{\pm}}^{fluxes}=H_{MIM2}+\int_{\Sigma}C_+\ .
\end{equation}

Hence, both Hamiltonians differ on a constant given by the value on $\int_{\Sigma}C_+$ and they have equal equations of motion. This relation  between these a priori unconnected two sectors sectors can be interpreted as a kind of  M2-brane 'duality'.

Furthermore, a third equivalence was found in \cite{mpdm:2019:flux}. It was shown that the supermembrane with fluxes $C_{\pm}$ compactified on $M_9\times T^2$ can be expressed as a $U(1)$ supermembrane on a twisted torus \cite{mpgm10Monodromy} whose connection is defined in terms of the monopole connection associated to the central charge and a new  dynamical $U(1)$ gauge field. It is constructed in terms of the  a multivalued $X_{hr}$ and a single-valued $A_{r}$ embedding maps,
\begin{equation}
\label{ec6notasseccion5}
\mathcal{A}=\frac{1}{2}\epsilon_{rs}(A^{r}dX_h^{s}-A^{s}dX_h^{r}+A^{r}dA^{s}) \ .    
\end{equation}

Understanding the dynamics of those scalar fields as well as $X^m$ is important also in the characterization of the  $U(1)$  gauge field and its
associated $U(1)$ curvature, $\mathcal{F}^{U(1)}=d\mathcal{A}$.  The symplectic curvature and the $U(1)$ curvature satisfy the very nontrivial property that $\mathcal{F}^{U(1)}=\mathcal{F}$ when expressed in terms of the embedding maps $X_{hr},A_{r}$. 
In this paper, for simplicity,  we will adopt the point of view of the characterization of the curvature in terms of the  symplectic gauge field $\mathbb{A}$ without further reference to the $U(1)$ gauge connection.


\section{General system of equations of motion}
\label{s3}
In this section we analyze the system of equations that represents the dynamics of the supermembrane on a quantized  constant background field $C_{\pm}$. In the following we will consider for simplicity the tension of the membrane $T=1$. The Lagrangian density of the theory can be obtained by  performing a usual Legendre transformation of the Hamiltonian density defined as \begin{equation}\mathcal{H}_{T}=\sqrt{W}\mathcal{H}_c+\Lambda \phi,\end{equation} where $\Lambda$ represents a Lagrange multiplier and $\phi=-\phi_{APD}$  represents the APD constraint given by equation (\ref{APD local}). It contains the  Lagrangian of MIM2, formerly obtained in \cite{Bellorin:Restuccia:2003}, plus a constant term associated to the flux contribution.  Putting together the central charge and the $C_+$ flux contribution in a single term $K=\frac{1}{8}n^2+nc_+$, the action is then,
\begin{equation}
\label{actionsuperM2centralchargecompleta}
\begin{aligned}
S=\int d\tau d\sigma^2\mathcal{L}_T=-\int d\tau d\sigma^2\sqrt{W}
\Big[\frac{1}{2}(D_i{X}_m)^2 +\frac{1}{4}\left\{X_{m}, X_{n}\right\}^{2}+\frac{1}{4}\mathcal{F}_{ij}^{2}\Big]+K.
\end{aligned}
\end{equation}
Since the $C_+$ contribution gives a constant term, that is added to the central charge contribution, its equations of motion are equal to those of the M2-brane with central charges. 
The symplectic field strength is defined as  $\mathcal{F}_{ij}=D_iA_j-D_jA_i+\{A_i,A_j\}$ and it now runs over the indices $i,j=0,r$ with $A_0=\Lambda$ and $D_0=\partial_{\tau}$. The symplectic covariant derivative gets also generalized $\mathcal{D}_iX^m=D_iX^m+\{A_i,X^m\}$.
As explained in \cite{Hoppe:PhdTesis}, \cite{Allen:Andersson:Restuccia:2010} the gauge freedom of the system allows to fix $\Lambda=0$ and then the Lagrangian density function $\mathcal{L}=\mathcal{L}_T-K$  reduces to 
\begin{equation}
\label{lagrangeanosuperM2centralchargecompleta}
\begin{aligned}
\sqrt{W}^{-1}\mathcal{L}=
\frac{1}{2}[(\dot{X}_m)^2 + (\dot{A}_{r})^2]-\frac{1}{4}\left\{X_{m}, X_{n}\right\}^{2}- \frac{1}{2}\left(\mathcal{D}_{r} X_m
\right)^2-\frac{1}{4}\mathcal{F}_{rs}^{2}
\end{aligned}
\end{equation}

The nonlinear system of equations that one has to solve is the following:
From  (\ref{lagrangeanosuperM2centralchargecompleta}) we derive the equations of motion for the dynamical fields $X^m(\tau,\sigma,\rho)$ and $A_{r}(\tau,\sigma,\rho)$
\begin{equation}
\label{equationofmotionnoncompact2}
\ddot{X}^m (\sigma, \rho, \tau)= \left\{X_{n}, \left\{X^{n}, X^{m}\right\}\right\}
+\left\{ {X}_{r},\mathcal{D}^{r}{X}^{m}\right\}\ ,
\end{equation}
\begin{equation}
\label{equationofmotionA4}
\begin{aligned}
\ddot{A}_{r}(\sigma, \rho, \tau) = \left\{ \mathcal{D}_{r} X^m, X_{m} \right\}  +\left\{  \mathcal{F}_{rs},{X}^{s} \right\} \ ,
\end{aligned}
\end{equation}
subject to satisfy
the APD constraint of the theory 
\begin{equation}
\label{constraint_A}
\mathcal{D}_{r} \Big(\dot{A}_r\Big)+\left\{X^m,\dot{X}^m\right\}=0\,\quad \text{for}\quad r=1,2,
\end{equation}
 and the topological central charge condition that restricts the winding numbers allowed for the harmonic contributions associated to the multivalued maps, $X_{rh}=R_{r}m_{r}^{s} \widehat{X}_s(\sigma,\rho)$,
 \begin{equation}\label{centralcharge1}
 n=det(m_{r}^{s}).
 \end{equation}
In order to find the admissible M2-brane solutions, the system of equations, (\ref{equationofmotionnoncompact2}), (\ref{equationofmotionA4}), (\ref{constraint_A}) and (\ref{centralcharge1}) must be solved.  For simplicity we will also  assume one of the scalar fields constant and we will fix the harmonic sector as in \cite{BRR2005nonperturbative}, as follows  
\begin{equation} \label{ansatz armonico} X_2=constant, \quad X_{h_{r}}=R_{r}(n_{r}\sigma+m_{r}\rho), \quad r=1,2.\end{equation} 
 For this ansatz $\sqrt{W}=1$. 
 There are some differences with the case analyzed in \cite{BRR2005nonperturbative}. The main one is associated to the topological restriction imposed called 'central charge condition'. Consequence of it is that appears a different degree of freedom, $A_{r}$. This field must be single-valued to define a physically admissible symplectic gauge connection $\mathbb{A}=dA_{r}$ in the theory. The E.O.M for $X_2$ constant, can be expressed in terms of complex variables $Z_a$ with $a=1,2,3$ where $Z_{a}(\tau , \sigma , \rho) = X_{2a+1}+iX_{2a+2}$. They  are, 
\begin{eqnarray}
\label{equationofmotionnoncompactcomplexZ}
\ddot{Z}_c&=& \frac{1}{2}\sum_{a=1}^{3}\left(\left\{\left\{Z_{c}, Z_{a}\right\},  \overline{Z}_{a}\right\}+\left\{\left\{Z_{c}, \overline{Z}_{a}\right\},  {Z}_{a}\right\}\right) 
+\sum_{r=9}^{10}\left\{X_{r},\mathcal{D}_{r} Z_c\right\},\\
\label{equationofmotionAcomplex}
\ddot{A}_{r}&=&
\frac{1}{2}\sum_{a=1}^{3}\left(\left\{\mathcal{D}_{r}Z_a,\overline{Z}_{a}\right\} +\left\{\mathcal{D}_{r}\overline{Z}_a,Z_{a}\right\}\right) + \sum_{s=9}^{10}\left\{\mathcal{F}_{rs},X^{s}\right\} \ ,
\end{eqnarray}
subject to the APD constraint of the theory: 
\begin{equation}
\label{constraint_complex_general}
\frac{1}{2}\sum_{c=1}^{3}\left(\left\{\dot{Z}_{c}, \bar{Z}_{c}\right\}+\left\{\dot{\bar{Z}}_{c}, Z_{c}\right\}\right) -\mathcal{D}_{r}\dot{A}_{r}=0 \ .
\end{equation}

%
 
Due to the complexity of the equations we will assume a separation of  the temporal and spatial dependence for the scalar field that represents the embedding into the noncompact sector.\\
We will assume

\begin{equation}
Z_a=f_a(\sig,\rho)e^{i\omega_a\tau}.
\end{equation}

\newpage
The E.O.M.  for $Z_a$ become 
{\small
\begin{eqnarray}
\label{EOM_Zc}
\omega_c^2 f_c=
&-&
\partial_{\sigma}^{2} f_c
\left[\sum_{r=9}^{10}
\left[R_{r} m_{r}+\partial_{\rho} A_{r}\right]^{2}
+\sum_{a=1}^3
(\partial_\rho f_a \partial_ \rho\bar f_{a})
\right]
\\
&+&\partial_{\sigma \rho}^{2} f_c
\left[\sum_{r=9}^{10}
2  \ \left(R_{r} n_{r}+\partial_{\sigma} A_{r}\right)  \left(R_{r} m_{r}+\partial_{\rho} A_{r}\right)
+\sum_{a=1}^3
\left[\partial_{\sigma} f_{a} \partial_{\rho} \bar{f}_{a}+h.{c}\right]
\right]
\nonumber\\
&-&\partial_{\rho}^{2}f_c 
\left[\sum_{r=9}^{10}
 \left(R_{r} n_{r}+\partial_{\sigma} A_{r}\right)^{2}
 +\sum_{a=1}^3
\left(\partial_ \sigma f_{a} \partial_ \sigma \bar f_{a}\right)
\right]
\nonumber\\
&-&\partial_{\sigma} f_c
\sum_{r,a}\left[
\partial_{\sigma \rho}^{2} A_{r}
\left( \partial_\rho A_{r} +R_{r} m_{r}\right)
-\partial_{\rho}^{2} A^ r
\left( \partial_\sigma A_{r} +R_{r} n_{r}\right)
+
\frac{1}{2}     \left[
(\partial_{\sigma \rho}^{2} f_{a} \partial_{\rho} \bar{f}_{a}
-
\partial_{\rho}^{2} f_a \partial_{\sigma} \bar{f}_{a})
+h.c
                \right]
\right]
\nonumber
\\
&-&\partial_{\rho} f_{c}
\sum_{r,a}\left[
\partial_{\sigma \rho}^{2} A_{r}
                \left( \partial_\sigma A_{r} +R_{r} n_{r}\right)
                -\partial_{\sigma}^{2} A^ r
                \left( \partial_\rho A_{r} +R_{r} m_{r}\right)
+
\frac{1}{2}     \left[
(\partial_{\sigma \rho}^{2} f_{a} \partial_{\sigma} \bar{f}_{a}
-
\partial_{\sigma}^{2} f_a \partial_{\rho} \bar{f}_{a})
+h.c
                \right]
\right]\ .
\nonumber
\end{eqnarray}
       }
       
The E.O.M for $A_{r}$ are:
{\small
\begin{eqnarray}\label{EOMForArZiagualfg}
\ddot{A}_{r}
&=&
\frac{1}{2}R_{r}\sum_{a=1}^{3}  \Big[
\left(n_{r} \partial^2_{\sigma\rho} f_{a} - m_{r} \partial_{\sigma}^{2} f_{a}\right) \partial_{\rho}\bar f_{a}
-
\left(n_{r} \partial_{\rho}^{2} f_{a} - m_{r} \partial_{\sigma \rho}^2  f_{a}\right) \partial_\sigma\bar f_{a} \Big]+ h.c\\
&+& \frac{1}{2} \sum_{a=1}^{3} \Big[
\partial_{\sigma}
\left( \partial_\sigma A_{r} \partial_{\rho} f_{a} - \partial_\rho A_{r} \partial_{\sigma} f_{a}\right)
\partial_{\rho} \bar f_{a} 
- 
\partial_{\rho}
\left( \partial_\sigma A_{r} \partial_{\rho} f_{a} - \partial_\rho A_{r} \partial_{\sigma} f_{a}\right) 
\partial_{\sigma}\bar f_{a}
\Big]+ h.c 
\nonumber \\
&+&\sum_{s=9}^{10}
\partial_\sigma
\Big[
R_{r}\left( n_{r} \partial_{\rho} A_{s}-m_{r} \partial_{\sigma} A_{s} \right)
-
R_s\left(n_s \partial_{\rho} A_{r}-m_s \partial_{\sigma}{A}_{r} \right)
\Big]
(\partial_\rho A_s+R_s m_s)
\nonumber
\\
&-&
\sum_{s=9}^{10}
\partial_{\rho}
\big[
R_{r}\left(n_{r} \partial_{\rho} A_{s}-m_{r} \partial_{\sigma} A_{s}\right)
-
R_s\left(n_s \partial_{\rho} A_{r}-m_s \partial_{\sigma}{A}_{r}\right)
\big]
(\partial_\sigma A_s+R_s n_s)
\nonumber\\
&+&
\sum_{s=9}^{10}
\partial_\sigma\left(\partial_{\sigma} A_{r} \partial_{\rho} A_{s}-\partial_{\rho} A_{r} \partial_{\sigma} A_{s}\right)
(\partial_\rho A_s+R_s m_s)
		\nonumber
		\\
&-&
\sum_{s=9}^{10}\partial_{\rho}\left(\partial_{\sigma} A_{r} \partial_{\rho} A_{s}-\partial_{\rho} A_{r} \partial_{\sigma} A_{s}\right)
(\partial_\sigma A_s+R_s n_s)\ ,\nonumber
\end{eqnarray}
        }
 and the APD constraint in components is
\begin{eqnarray}
\label{constraint_f_g_general_for_Z}
0&=&
{i\omega}
\sum_{c=1}^{3} 
(\partial_\sigma f_{c} \partial_\rho \bar{f}_{c} -  \partial_\rho {f}_{c} \partial_\sigma \bar{f}_{c})
 +\sum_{r=9}^{10}R_{r}(\partial_\sigma \dot{A}_{r} m_{r} -  \partial_\rho \dot{A}_{r} n_{r})\\&&+\sum_{r=9}^{10}(\partial_\sigma \dot{A}_{r} \partial_\rho A_{r} -  \partial_\rho \dot{A}_{r} \partial_\sigma A_{r}).\nonumber
\end{eqnarray}

The system of equations is subject to the topological restriction given by the non-vanishing central charge condition 
\begin{equation}\label{centralcharge}
    n_{9}m_{10}-n_{10}m_{9}=n\ne0 \quad n\in \mathbb{Z}  ,
\end{equation}
 where mixed  partial derivatives independent have been assumed to be independent of the derivation order $\partial^2_{\sigma\rho}=\partial^2_{\rho\sigma}$.
One can realize that the system of equations is highly nonlinear. We will assume in the following different types of ansatz for the complex scalar fields $Z_a$ and for $A_{r}$. We analyze different cases for the complex scalar field: i) $Z_a$ constant, ii) a $Z_a$ with $f_a=r_ae^{i(k_a\sigma+l_a\rho_a)}$ and constant $r_a$. We will denote it indistinctly as rotating or spinning. This ansatz was originally proposed by  \cite{FloratosRotatingToroidalB2002ranes, BRR2005nonperturbative}. iii) $Z_a$ with $f_a(\sigma,\rho)$ an arbitrary real function. We will denote this last ansatz as 'Q-ball like' in spite of the  toroidal symmetry associated to the membrane worldvolume. The reason for this name  is that this type of ansatz has been used in the context of Q-balls, and the fact that there is an associated $U(1)$ Noether charge defined as 
\begin{equation}
Q_a=i\int_{\Sigma} d\rho d\sigma (\overline{Z_a}\dot{Z_a}-\dot{\overline{Z_a}}Z_a),
\end{equation}
which for the ansatz proposed corresponds to 
\begin{equation}\label{Q-ball ansatz}
Q_a=\omega_a\int d^2\sigma f^2_a(\sigma,\rho).
\end{equation}
In any case, its solitonic nature, if it exists requires a more profound study outside of the scope of this paper. Here, it represents just a name classifying the type of ansatz. In the following we will shorten it as QBL ansatz.

On the gauge field side, we will consider  six types of different embeddings associated to $A_{r}$: a constant one, two types of polynomial ansätze that we denote as linear and 'separable', a periodic regular sinusoidal solution, a rotating ansatz, and a QBL ansatz. We will see that the APD constraint as well as the central charge condition eliminates most of the possible embeddings.
\section{Spinning membrane solutions of the M2-brane with fluxes \texorpdfstring{$C_{\pm}$}{}} 
\label{s4}
In this section we show that spinning embeddings are solutions to the M2-brane with fluxes $C_{\pm}$. We will also show that the mass operator of the M2-brane with fluxes $C_{\pm}$ contains the energy contribution obtained in \cite{BRR2005nonperturbative} and hence the subset of the spinning solutions found in the aforementioned paper that also preserve the worldvolume flux condition, i.e. the central charge condition as also solutions of the M2-brane with $C_{\pm}$ fluxes. 
In order to obtain spinning solutions of the M2-brane  we fix the background component of the three-form $C_+=0$ imposing a non-vanishing flux condition over $C_-$ and we assume the following embedding:
\begin{subequations}\label{AnsatzAnalogoCFlujo}
\begin{alignat}{3}
&A_{r}(\tau, \sigma, \rho)= constant,\quad  
X_2(\tau, \sigma, \rho) = constant \ , \\
&Z_{a}(\tau, \sigma, \rho)=  r_{a} e^{i (k_a\sigma+l_a\rho)}e^{i\omega_a \tau}, \quad \textrm{with}\quad  a=1, 2, 3;
\\
&X_{r}(\tau, \sigma, \rho) = R_{r}\left(n_{r} \sigma+m_{r} \rho\right)+q_{r} \tau + A_{r}(\tau, \sigma, \rho), \qquad \quad r=9, 10 \ ,
\end{alignat}
\end{subequations}
where here it is assumed $r_a$  to be constant,  $\omega_a$ a rotation frequency, $k_a, l_a$ integers associated to the Fourier modes and $q_{r}$ integers parametrizing the KK modes.
 The E.O.M. system particularized to $A_{r}=constant$ is the following one. For $Z_c$:

{\small
\begin{eqnarray}
\label{frecuencia_BRR}
\omega_c^2 f_c=
\sum_{r=9}^{10}\sum_{a=1}^3\Bigg[
&-&
\partial_{\sigma}^{2} f_c
\left[
\left(R_{r} m_{r}\right)^{2}
+
(\partial_\rho f_a \partial_ \rho\bar f_{a})
\right]
-
\frac{1}{2} 
\partial_{\sigma} f_c
                \left[
(\partial_{\sigma \rho}^{2} f_{a} \partial_{\rho} \bar{f}_{a}
-
\partial_{\rho}^{2} f_a \partial_{\sigma} \bar{f}_{a})
+h.c
               \right]
\nonumber
\\
&+&\partial_{\sigma \rho}^{2} f_c
\left[
2  \ \left(R_{r}^2 n_{r} m_{r}\right) 
+
(\partial_{\sigma} f_{a} \partial_{\rho} \bar{f}_{a}+h.{c})
\right]
\\
&-&\partial_{\rho}^{2}f_c 
\left[
 \left(R_{r} n_{r}\right)^{2}
 +
\left(\partial_ \sigma f_{a} \partial_ \sigma \bar f_{a}\right)
\right]
-
\frac{1}{2}
\partial_{\rho} f_{c}
\left[
(\partial_{\sigma \rho}^{2} f_{a} \partial_{\sigma} \bar{f}_{a}
-
\partial_{\sigma}^{2} f_a \partial_{\rho} \bar{f}_{a})
+h.c
\right]
                    \Big]\ .
\nonumber
\end{eqnarray}
}

The E.O.M for $A_{r}$, {(\ref{EOMForArZiagualfg})} are:

\begin{equation}\label{EOMForArconArConst}
    0=\sum_{a=1}^{3}  \Big[
[\left(n_{r} \partial^2_{\sigma\rho} f_{a} - m_{r} \partial_{\sigma}^{2} f_{a}\right) \partial_{\rho}\bar f_{a}
-
\left(n_{r} \partial_{\rho}^{2} f_{a} - m_{r} \partial_{\sigma \rho}^2  f_{a}\right) \partial_\sigma\bar f_{a}] + h.c \Big]
\end{equation}

See that even for $A_{r}$ constant, the above equation is not trivial.
The APD constraint in components becomes
\begin{eqnarray}
\label{constraint_ArConst}
0&=&
\sum_{c=1}^{3} 
\left[(\partial_\sigma f_{c} \partial_\rho \bar{f}_{c} -  \partial_\rho {f}_{c} \partial_\sigma \bar{f}_{c})\right]. \ 
\end{eqnarray}
The central charge condition (\ref{centralcharge}) must also be satisfied. If now  the ansatz (\ref{AnsatzAnalogoCFlujo}) is substituted, it is possible to solve all of the equations and constraints and
one obtains a spinning $Z_a$ solution with a frequency explicitly given by 
\begin{eqnarray}
\label{eomwithansatz}
\omega_{c}^{2}=
\sum_{a=1}^{3} r_{a}^{2}  \left(k_{a} l_{c}-l_{a} k_{c}\right)^{2}+
\sum_{r=9,10} R_{r}^{2} \left(n_{r} l_{c}-m_{r} k_{c}\right)^{2},\  
\end{eqnarray}
for $l_c,k_c$ arbitrary and $m_s,n_s$ satisfying the central charge condition. 
These results are in complete agreement with the results obtained in \cite{BRR2005nonperturbative} restricted to satisfy in addition the central charge condition. 

In the following, we will show that the mass operator of the M2-brane with fluxes $C_{\pm}$ formulated on $M_9\times T^2$ target space contains the energy expression obtained in \cite{BRR2005nonperturbative} and hence, it is natural to explain that all of their spinning solutions, that we denote succinctly by BRR solutions, that also satisfy the central charge condition are also solutions of the M2-brane with  $C_{\pm}$ fluxes, as we have previously shown by direct computation.
 By using the mass operator of MIM2 \cite{GMPR2012}, and the duality reviewed in section 2., it is straightforward to obtain the mass operator for the M2-brane with fluxes,
\begin{equation}\label{EnergiaMPilar}
\mathcal{M}^2= T^2 [(2 \pi R)^2 n \, (Im \tilde\tau)] ^2 
+
\left(\frac{m|q \tilde\tau-p|}{R [I_{m}\tilde\tau]}\right) ^2
+
T^{2/3}H_{MIM2}+\int_{\Sigma}C_+ \ ,
\end{equation}
where  $T$ denotes the membrane tension, $n$ the worldvolume flux units or equivalently the central charge, $m$ is a relative prime integer number, $q,p$ are integers denoting the KK charges, $R$ the radius and $\widetilde{\tau}=\widetilde{\tau}_1+i\widetilde{\tau}_2$ the Teichmuller parameter of the target space 2-torus. 

The first two terms in the mass operator correspond respectively to the central charge and the Kaluza Klein (KK) momentum contribution.
In order to reproduce the BRR energy expression $\mathcal{M}_{BRR}$ from the M2-brane with fluxes theory in the LCG, we fix the background component of the three-form $C_+=0$ imposing a non-vanishing flux condition over $C_-$ and we assume $A_{r}$ to be constant with $Z_a$ and $X_{r}$ satisfying the embedding ansatz given by (\ref{AnsatzAnalogoCFlujo}).
If we assume a rectangular 2-torus, i.e. $\tau_1=0$, and we define new KK integers as $\tilde{n}_{10}=m p, \tilde{n}_{9}=m q$, the KK momentum contribution for this ansatz can be re-expressed as
\begin{equation}
\left(\frac{m|q \tilde\tau-p|}{R_{11} [I_{m}\tilde\tau]}\right)=\frac{\tilde{n}_{9}^{2}}{R_{9}^{2}}+\frac{\tilde{n}_{10}^{2}}{R_{10}^{2}}=(P_9^{kk})^2+(P_{10}^{kk})^2, \qquad \text{     $\tilde{n}_{10}=m p, \ \ \tilde{n}_{9}=m q$}\ ,
\end{equation}
being $R_{9} = R_{11}$, and $R_{10}=R_{11} Im \tilde\tau$. 
The central charge contribution, particularized to this background becomes 
\begin{eqnarray}
T^2 [(2 \pi R_{11})^2 n \, (Im \tilde\tau)] ^2 &=&T^{2} 4 \pi^{2} R_{9}^{2} R_{10}^{2}\left(n_{9} m_{10}-m_{10} m _9\right)\ ,
\end{eqnarray} 
where we have used the central charge expression (\ref{centralcharge}) corresponding to the determinant of the wrapping numbers.

Finally, in \cite{BRR2005nonperturbative}  the angular momentum defined  was defined in terms of the frequency modes as 
\begin{equation}
J_{a}=\int d\sigma\rho \frac{\delta S}{\delta\dot{\beta}^a}=4 \pi^{2} T r_{a}^{2} \omega_{a}\ .
\end{equation}
Substituting it in the Hamiltonian and using the equation of motion associated to the complexified embedding maps $Z_a$ and the value for the frequency $\omega_c$ whose expression is (\ref{frecuencia_BRR}), it is then possible to formulate the mass operator on this background,
\begin{equation}\label{EnergiaRusso}
E= 2(4 \pi^2 T) ^{2/3} J_a\omega_a + (4 \pi^2 T) ^{2}R_9^2 R_{10}^2(n_9m_{10}-n_{10}m_9)^2+ \frac{\tilde{n}_9^2}{R_9^2}+\frac{\tilde{n}_{10}^2}{R_{9}^2} \ .
\end{equation}
It corresponds to the energy operator of a spinning membrane obtained in \cite{BRR2005nonperturbative}.
Since the M2-brane with fluxes is formulated in the LCG, one plane less is observed. We should recall that there is an extra topological condition imposed on the M2-brane with fluxes that is not present in the aforementioned formulation. It restricts the set of allowed spinning solutions of \cite{BRR2005nonperturbative} to the subset that also satisfy it.  
 In summary, we have shown that the BRR results can be obtained from the M2-brane with fluxes once  the background is fixed, i.e. $C_+=0$  and we have frozen the dynamical degree of freedom associated to the gauge symplectic connection $A_{r}$ and the ansatz (\ref{AnsatzAnalogoCFlujo}) is assumed. This background has also restricted the APD constraint expression to the one used by \cite{BRR2005nonperturbative}. Hence BRR spinning solutions that also satisfy the $C_-$ flux condition (\ref{cargaCentral})  are naturally contained in the allowed spinning solutions of the M2-brane with $C_{\pm}$ fluxes.

\section {New solutions with constant gauge field} 
\label{s5}
In this section we obtain approximate solutions to the E.O.M using a QBL ansatz on $Z_a$. These type of solutions had not been considered previously on this target space, and they are interesting since in principle they could model non-topological solitons. However at this level, we do not analyze any solitonic behaviour but we just focus on the admissible solutions that model the dynamics of the M2-brane with fluxes. The QBL ansatz corresponds to $Z_a=f_a(\sigma,\rho)e^{i\omega_a\tau}, a=1,2,3$ with $f_a(\sigma,\rho)$ a real function. The set of equations of motion with $A_{r}= constant$ simplifies to
\begin{eqnarray}\label{EOMZeqfgAConst}
 -\omega_{c}^{2}f_c=
& & \sum_{a=1}^3\left[ 
(\partial_\sigma^2 f_{c}\partial_\rho f_{a} - 2\partial^2_{\sigma\rho} f_c \partial_\sigma f_a + \partial_\sigma f_c\partial^2_{\sigma\rho} f_a - \partial_\rho f_c \partial_\sigma^2 f_a \ )\right. \partial_\rho {f}_a 
\\
& &\quad\ \  +\left(\partial_{\rho}^{2} f_{c} \partial_{\sigma} f_{a}-\partial_{\sigma} f_{c} \partial_{\rho}^{2} f_{a}+\partial_{\rho} f_{c} \partial^2_{\rho\sigma} f_{a}\right) \partial_\sigma  f_a] \nonumber
\\
&+&\sum_{r=9,10}
R^2_{r}\left(m_{r}^2 \partial_{\sigma}^{2}+n_{r}^2 \partial_{\rho}^{2}-2n_{r}m_{r}\partial_{\rho\sigma}^2\right) f_{c}
\nonumber
\end{eqnarray}

The E.O.M for $A_{r}$ are nontrivial in spite that $A_{r}$ is constant, indeed they represent a restriction to the solutions to (\ref{EOMZeqfgAConst}):

\begin{eqnarray}\label{EOMForArZiagualfgArConst}
0= \sum_{a=1}^{3}  \left[
\left(m_{r} \partial_{\sigma}^{2} f_{a}-n_{r} \partial^2_{\sigma \rho} f_{a}\right) \partial_\rho f_{a}
-
\left(m_{r} \partial^2_{\rho \sigma} f_{a}-n_{r} \partial_{\rho}^{2} f_{a} \right)\partial_{\sigma} f_{a}  \right]\ .
\end{eqnarray}

The wrapping numbers must also satisfy to the central charge condition (\ref{centralcharge}). The APD constraint associated to the complex scalar field $Z_a$ verifies identically leaving only a residual constraint over the $A_{r}$ maps that is  trivially satisfied for $A_{r}$ constant. 

Since the system of equations is highly non-linear in order to simplify it further we  perform an approximation that we denote as 'small Q-ball'. The approximate $Z_a$ will be parametrized by an arbitrary constant $\lambda$ that we will assume to be small with
 \begin{equation}
 Z_a=\lambda f_a(\sigma,\rho) \, e^{i\omega\tau}\ .
 \end{equation}
We neglect the terms of order $O(\lambda^2)$. See that this approximation acts on the terms that are products of $f_a$ but it does not affect to the order of derivatives appearing in the equation. By doing it  the equation (\ref{EOMForArZiagualfgArConst}) is not considered and (\ref{EOMZeqfgAConst}) gets simplified. 

We obtain an infinite set of solutions with a discrete set value of frequencies allowed. Indeed the membrane eigenfunctions can be determined numerically and they  have an associated 'breathing mode' determined by the frequency eigenvalue. This is a very interesting result associated to the fact that we obtain an elliptic operator\footnote{For $\frac{1}{4}C_2^2-C_1C_3<0$ the equation is elliptic. More precisely, the previous condition is always satisfied since $-R_9^2R_{10}^2 (m_9n_{10}-m_{10}n_9)^2<0$ for $n\neq 0$ and any arbitrary value of the constants $R_s,m_s,n_s$. For vanishing central charge,   $n=0$ and $\frac{1}{4}C_2^2-C_1C_3=0$, then the operator is parabolic.}, that it admits an infinite discrete set of solutions. 

 Now we consider the excitations of the M2-brane in the three complex noncompact planes equal, i.e. that $Z_a=Z, \forall a$. The motion of each plane disentangles and hence many of the terms  in the equation (\ref{EOMZeqfgAConst}) disappear. Indeed the approximation only acts eliminating the restriction imposed by the  $A_{r}$  EOM in (\ref{EOMForArZiagualfgArConst}) and it leaves untouched the $Z_a$ EOM.
 
  In this isotropic regime, the $Z_a$ E.O.M for each $a$ reduce to

\begin{equation}
\label{qballaproximation}
-C_{1}\left(\partial_{\sigma} ^2 f\right)+C_{2}\left(\partial^2_{\sigma \rho} f\right)-C_{3} \partial_{\rho}^2 f=\omega^{2}f \ ,
\end{equation}

with
\begin{eqnarray}
    C_1= \sum_{r} R_{r}^2m_{r}^2,\quad
    C_2= 2\sum_{r} R_{r}^2m_{r}n_{r},\quad
    C_3=  \sum_{r} R_{r}^2n_{r}^2\ .
\end{eqnarray}

 The APD constraint is trivially verified and the central charge condition (\ref{centralcharge}) must be satisfied.

\subsection{Numerical approach to the solution}
A numerical approximation to the solution of the equation  (\ref{qballaproximation}), can start by representing this equation as an eigenvalue problem,

\begin{equation}
{\cal L}_{p} f_{\alpha}(\sigma,\rho) = \omega^2 ~ f_{\alpha}(\sigma,\rho),
\label{Eigenvalue-eq}\ 
\end{equation}

\noindent
 using the corresponding eigenfunctions and eigenvalues to represent any particular solution.

With this in mind we have defined ${\cal L}_{p}$, as the differential operator in the left side of the equation (\ref{qballaproximation}), $p = \{C_1,C_2, C_3\}$ the parameters associated to these operator and $\omega$ the set of eigenvalues associated to eigenfunctions. For simplicity, we will focus on a single plane. 

In order to estimate the numerical solutions, it is necessary to establish boundary conditions on the boundary ($\Gamma$) of the domain of $f$. In this case we will present three types of these conditions: i) Periodic conditions, ii) Periodic conditions with restrictions and iii) Dirichlet conditions. 

In all cases, the parameters $ C_1 $, $ C_2 $, and $ C_3$ were chosen in order to satisfy the central charge condition (\ref{centralcharge}) with $ n_k, m_k, R_k \in [1,10^2] $ and $k=9, 10$. In this case, there are $10^{12}$ $3$-tuples $\{C_1, C_2, C_3\}$.

\begin{figure}[ht]
	\begin{center}
		\includegraphics[width=0.55\textwidth]{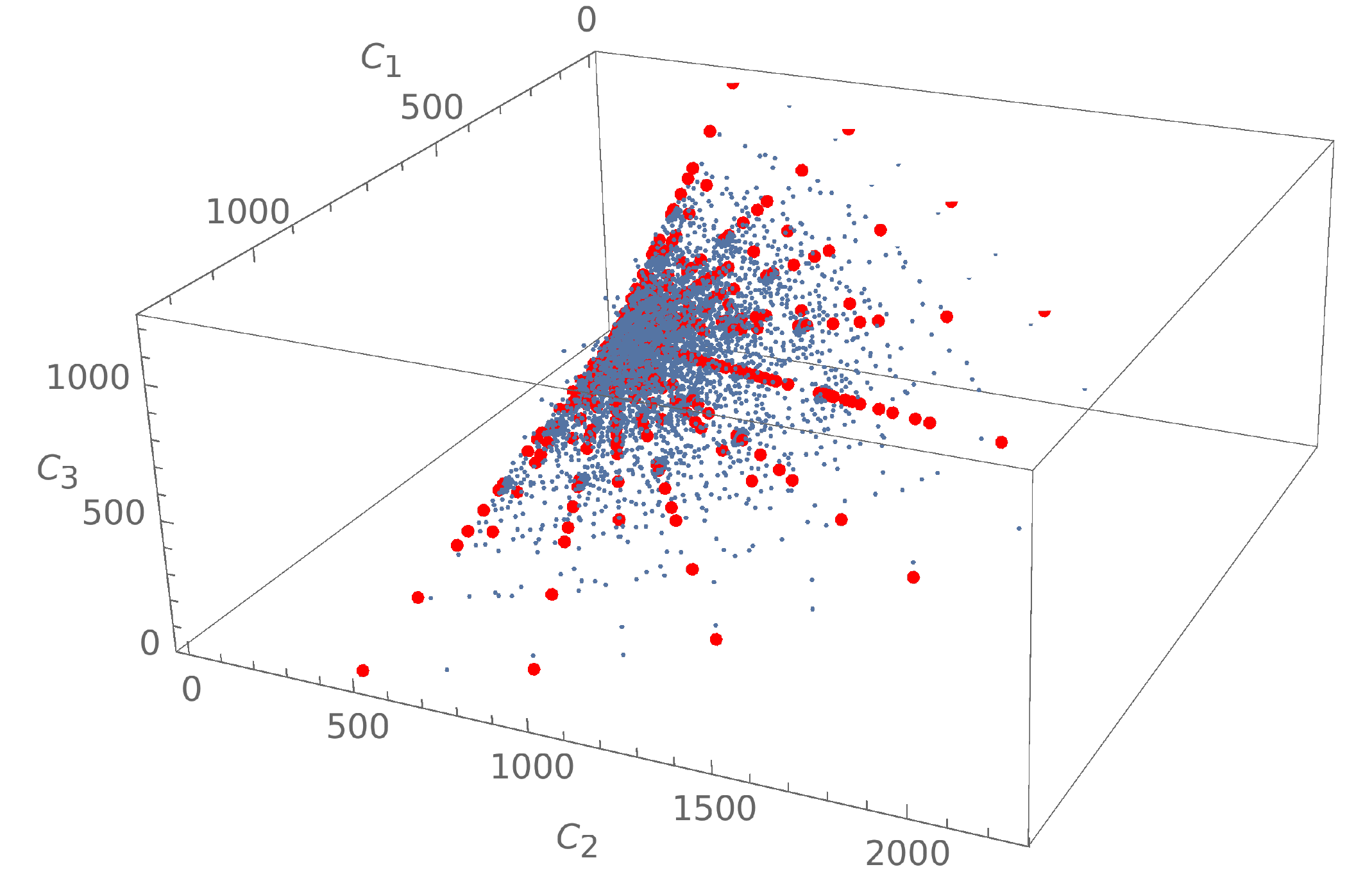}
	\end{center}
\caption{Distribution of the parameters of the differential operator. Grey points parametrize the $3$-tuples associated to the central charge condition and red points those with vanishing central charge.}
\label{Parameters}
\end{figure}

The figure \ref{Parameters} shows an interesting structure in the distribution of the first $5^6$ of these $3$-tuples (gray dots) and how the $3$-tuples that do not meet the central charge condition (\ref{centralcharge}) (big red dots) are inserted into that structure. Although in this case we use a small subset of the possible values of the parameters, our numerical experiments show that the structure of their distributions is maintained in larger sets of the order of $10^6$  data points. The distribution of the allowed discrete solutions in the set of parameters for the elliptic operator, i.e. its landscape, clearly dominates over those with zero central charge.

It is interesting to realize that the set solutions associated to the parabolic operator describing the solutions of the bosonic rotating membranes trivially wrapped follows a radial structure. 
\subsubsection{Periodic boundary conditions} These are the natural boundary conditions for the toroidal bosonic membrane.
Here we set:
\begin{eqnarray}
f(0,\rho) = f(2 \pi, \rho), \qquad
f(\sigma,0) = f(\sigma,2 \pi)\ ,
\end{eqnarray}

\noindent
with $n_9=62$, $n_{10}=28$, $m_9=70$, $m_{10}=73$, $R_9=4$ and $R_{10}=3$. In this case the first $9$  eigenvalues ($\omega/C_1's$) are:
$5.74486 \times 10^{-16}$, $0.152354$, $0.152354$, $0.39028$, $0.390286$, $0.54258$, $0.54258$, $0.547774$, $0.547774$.

\begin{figure}[ht]
\begin{center}
		\includegraphics[width=0.9\textwidth]{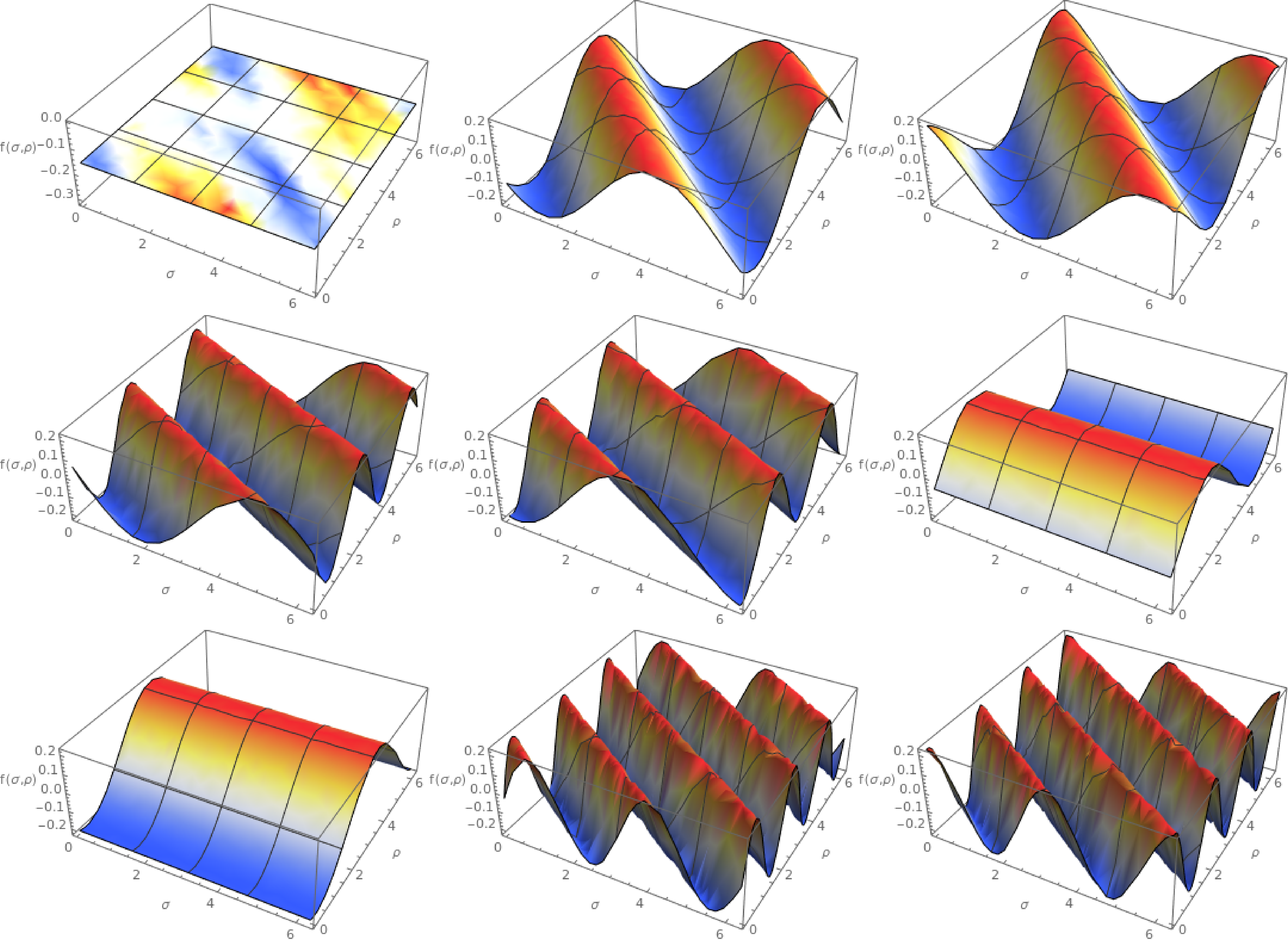}
	\end{center}
\caption{First nine eigenfunctions of the differential operator for periodic boundary conditions.}
\end{figure}
The eigenvalues are degenerated for the set chosen except the first one. The degeneration seems to be removed for larger values of the set of parameters. Although this is not the case for the particular example considered, see that the central charge can be kept small for arbitrarily large wrapping numbers. 
\subsubsection{Periodic boundary conditions with restrictions}
The periodic boundary conditions for the toroidal bosonic membrane can admit restrictions still satisfying the equations. In this case one of the points satisfies Dirichlet conditions. It is assumed to be fixed to zero. 
Here we set:
\begin{eqnarray}
f(0,\rho) = f(2 \pi, \rho),  \quad
f(\sigma,0)= f(\sigma,2 \pi),  \quad
f(0,0) &=& 0,
\end{eqnarray}

\noindent
with $n_9=62$, $n_{10}=28$, $m_9=70$, $m_{10}=73$, $R_9=4$ and $R_{10}=3$. In this case the first $9$  eigenvalues ($\omega/C_1's$) are: 
$0.012460$, $0.152354$, $0.180237$, $0.390285$, $0.413282$, $0.542579$, $0.545055$, $0.547773$, $0.581044$. In this case the degeneration of the eigenvalues is removed. 

\begin{figure}[ht]
	\begin{center}
		\includegraphics[width=0.9\textwidth]{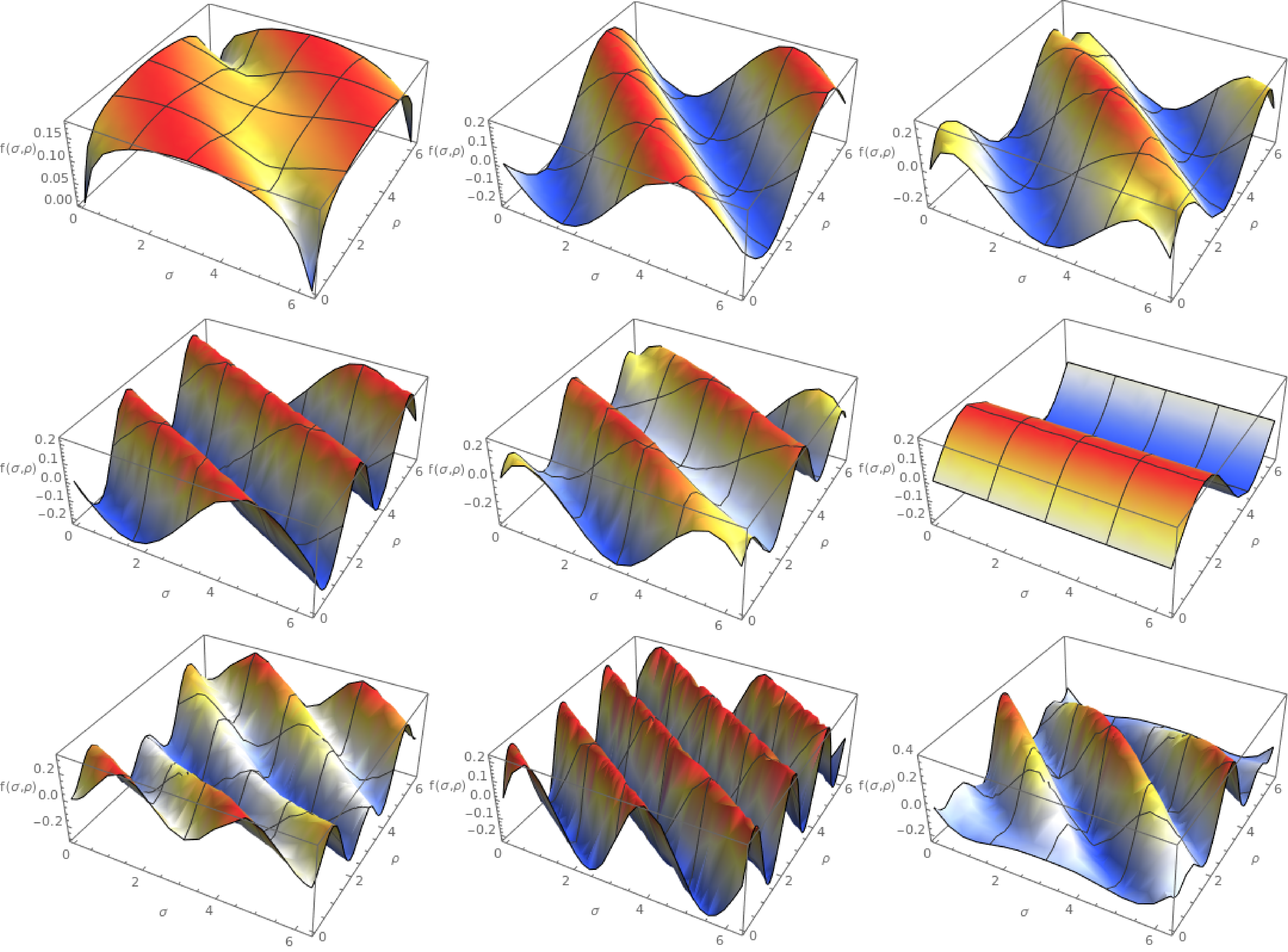}
	\end{center}
\caption{First nine eigenfunctions of the differential operator for periodic boundary conditions with restrictions.}
\end{figure}

\subsubsection{Dirichlet conditions}  This last case corresponds to well-known boundary conditions. It can be naturally imposed on open membranes. The closed membrane would not admit this type of boundary conditions to produce nontrivial solutions. Here we only use it for illustrative purposes. We can see the change in the allowed eigenfunctions with respect to the previous cases analyzed due to the effect of the boundary conditions. In particular we can observe the effect in the behaviour comparing with respect to the restricted case.  Here we set:
\begin{eqnarray}
f(0,\rho) = 0, \quad
f(\sigma,0) = 0,  \quad
f(2 \pi,\rho) = 0,  \quad
f(\sigma,2 \pi) = 0,
\end{eqnarray}
%
with $n_9=62$, $n_{10}=28$, $m_9=70$, $m_{10}=73$, $R_9 = 4$ and $R_{10} = 3$. In this case the first $9$  eigenvalues ($\omega/C_1's$) are: 
$0.289021$, $0.390560$, $0.516266$, $0.643088$, $0.781355$, $0.931779$, $1.061183$, $1.090063$, $1.229781$. In this case the eigenvalues are not degenerate. 

\begin{figure}[ht]
	\begin{center}
		\includegraphics[width=0.9\textwidth]{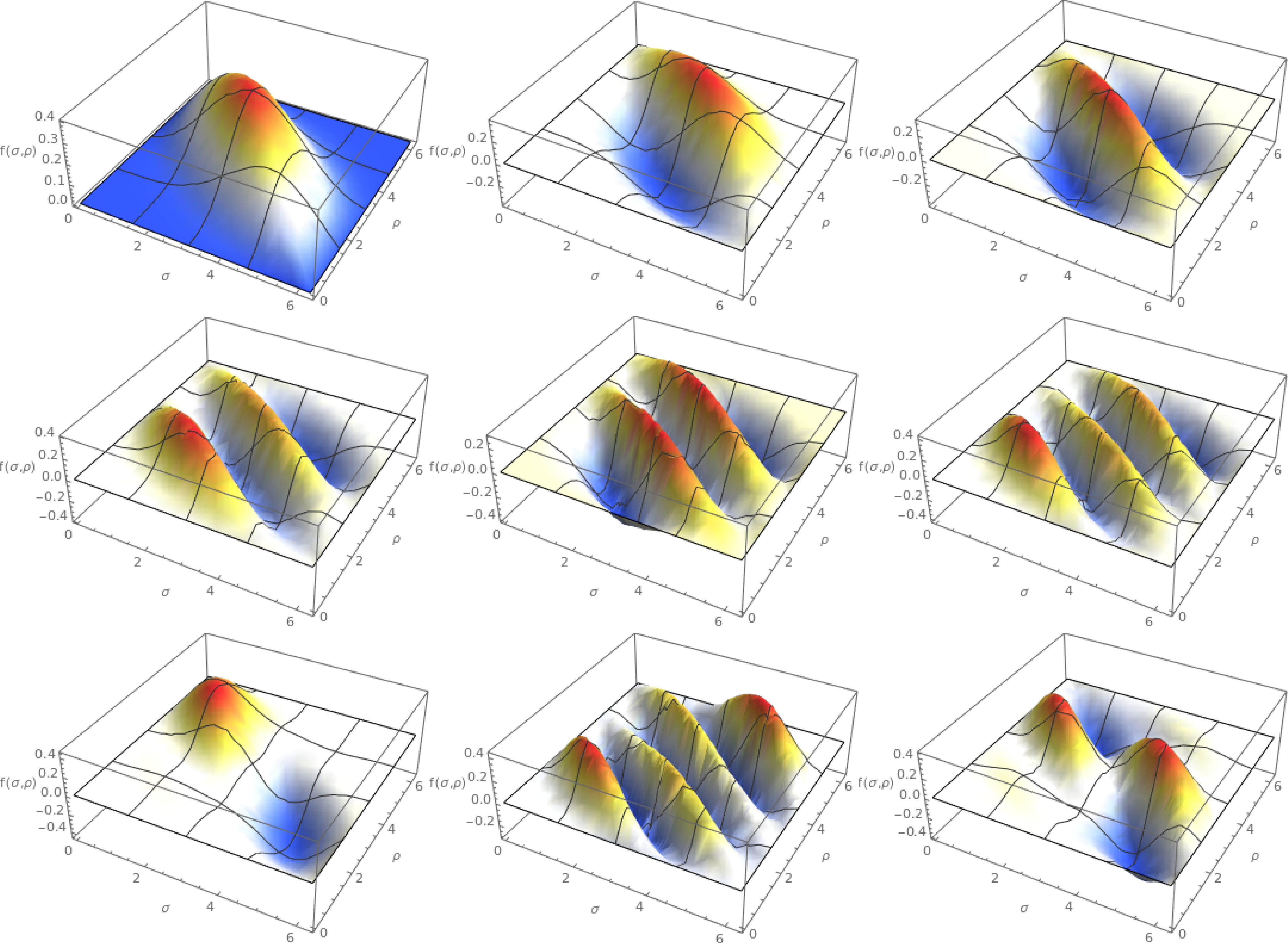}
	\end{center}
\caption{First nine eigenfunctions of the differential operator for  Dirichlet boundary conditions.}
\end{figure}


 Summarizing, the bosonic M2-brane with fluxes $C_{\pm}$ admits an approximate QBL solution propagating on the target space. 
 The background fluxes  and hence the central charge condition impose an important restriction on the equations such that even in the case considered here, with no propagating symplectic gauge field, one must take into account the  multivalued constant maps $X_{hr}$ contributions. However,  these results  does not exclude in any way the possibility to obtain an exact QBL solution for the complete system of equations of the M2-brane with $C_{\pm}$ fluxes can exist, but its study is out of the scope of the present work. 

\section{ Approximate solutions with a dynamical gauge field}
\label{s6}
The M2-brane with fluxes possesses a symplectic gauge field, an aspect that had not been analyzed in the preceding sections. In this section we will discuss approximate solutions to the problem. Firstly we search for solutions when an approximation is imposed on both dynamical fields $Z_a, A_{r}$, secondly we study the case in which the approximation is only imposed to the $Z_a$ and thirdly the case in which the approximation is only imposed on the $A_{r}$.
We assume approximate QBL and approximate spinning embeddings on $Z_a$ and we analyze the allowed field configurations for $A_{r}$. 
\subsection{First order approximation on \texorpdfstring{$Z_a$}{} and \texorpdfstring{$A_{r}$}{}}
We are interested in characterize further the QBL solutions for a bosonic M2-brane with  $C_{\pm}$ fluxes when a nontrivial $A_{r}$ dependence is included. As previously considered in section 5. we perform an approximation that simplifies the EOM. In this opportunity the approximation will be done on the complex scalar field $Z_a$ and on the $A_{r}$. We impose the same approximation to the one performed in section 5. In particular, if $Z_a=\lambda \widetilde{Z_a}$ and $A_{r}=\alpha \widetilde{A}_{r}$ such that  $O(\alpha)\sim O(\lambda)$,  we will neglect $O(\lambda^2)$, $O(\alpha^2)$ and $O(\alpha.\lambda)$.   One can impose the same approximation on both dynamical fields, the complex scalar field and the $A_{r}$ global functions associated to the symplectic gauge field. The  equations of motion for  $Z_a=\lambda f_a(\sigma,\rho) \, e^{i\omega\tau}$  (\ref{EOM_Zc})  in the approximation become
\begin{eqnarray}\label{EOMZAproxAprox}
{\omega_c^2 f_c}
&=&
-
D_1\ \partial_\sigma^2 f_c
+
D_2\ \partial^2_{\sigma\rho}  f_c
- D_3\ \partial_\rho^2 f_c .
\end{eqnarray}


In the approximation the E.O.M for the $A_{r}$ become simplified

\begin{eqnarray}\label{EOMAADobleAprox}
\ddot{A}_{r}=
&-& 
\sum_{s=9}^{10}\left[
 2 R_{s}^2 m_{s}  n_{s} \partial^2_{\sigma \rho} A_{r}-\left(R_{s} m_{s}\right)^{2} \partial_{\sigma}^{2} A_{r}-\left(R_{s} n_{s}\right)^{2} \partial_{\rho}^{2} A_{r}
\right]\\
&-& 
\sum_{s=9}^{10}R_{r}R_{s}\left[
 m_{s}m_{r} \partial_{\sigma}^{2} A_{s}
+  n_{s} n_{r} \partial_{\rho}^{2} A_{s}
-\left( m_{s} n_{r}+ n_{s} m_{r}\right) \partial^2_{\sigma \rho} A_{s}
\right].\ 
\nonumber
\end{eqnarray}

With respect to the APD first class constraint of the theory,  the part of the APD constraint associated to $Z_a$ is automatically satisfied and there is a residual condition over the $A_{r}$ contribution that still has to be satisfied,
\begin{eqnarray}
\label{constraint_f_g_general_for_Z_aprox_2}
&&\sum_{r=9}^{10}R_{r}(\partial_\sigma \dot{A}_{r} m_{r} -  \partial_\rho \dot{A}_{r} n_{r})=0\ .
\end{eqnarray}
Considering the above system of  equations (\ref{EOMZAproxAprox})
 (\ref{EOMAADobleAprox}) and (\ref{constraint_f_g_general_for_Z_aprox_2})  and assuming an admissible periodic regular solution for $A_{r}$ given by,
 \begin{equation}\label{sinusoidal}
     A_{r}=a_{r}\sin{\sigma}+b_{r}\sin{\rho}+c_{r}\tau\ .
 \end{equation}
 The equations (\ref{EOMAADobleAprox}) and (\ref{constraint_f_g_general_for_Z_aprox_2}) are satisfied for
\begin{equation}\label{coeficientes}
m_{10}=\frac{R_{9}}{R_{10}}\left(\frac{a_{10}}{a_{9}}\right) m_{9}
\quad;\quad
n_{10}=\frac{R_{9}}{R_{10}}\left(\frac{b_{10}}{b_{9}}\right) n_{9}\ .
\end{equation}
This implies that $(\frac{R_{9}a_{10}}{R_{10}a_{9}})$ and $(\frac{R_{9}b_{10}}{R_{10}b_{9}})$ must be an integer.
 The central charge is given by the following expression
\begin{equation}
n=\left(\frac{a_{10}}{a_{9}}-\frac{b_{10}}{b_{9}}\right) \frac{R_{9}}{R_{10}} n_{9} m_{9}\ ,
\end{equation}
subject to $ \frac{a_{10}}{a_{9}}\neq\frac{b_{10}}{b_{9}}$ to ensure being different from zero. The $A_{r}$ function coefficients become restricted to satisfy the equations. Indeed $a_{r}$ and $b_{r}$ become proportional to each other but with a proportionality constant different form one and such that its product with the radii ratio gives an integer. The equation (\ref{EOMZAproxAprox}) now becomes expressed in terms with the coefficients given by 
            \begin{eqnarray}\label{coeficientes D}
    D_1&=& R_9^2m_9^2\left[1+\left( \frac{a_9}{a_{10}}\right)^2\right],
 \nonumber   \\
    D_2&=& 2R_9^2m_9n_9\left[1+\left( \frac{a_9}{a_{10}}\right)\left( \frac{b_9}{b_{10}}\right)\right],
  \nonumber   \\
    D_3&=&  R_9^2n_9^2\left[1+\left( \frac{b_9}{b_{10}}\right)^2\right].
            \end{eqnarray}
            
In principle these equations allow general embeddings  with $f_a$ complex or real that include  spinning solutions, QBL solutions or spinning QBL solutions.  In the following we analyze two different cases: in the first one we consider $Z_a$ being described by a  QBL ansatz on the three complex planes. This implies taking the $f_a$ real.  In the second one $Z_a$ is describes by a spinning M2-brane on those three planes. For that case, the $f_a$ are complex functions.

 \paragraph{Approximate QBL solutions.} The equation (\ref{EOMZAproxAprox}) for the $Z_a$ approximate QBL embedding $Z_a=f_a(\sigma,\rho)e^{i\omega_a\tau}, a= 1,2,3$ with $f_a(\sigma,\rho)$ a real function. As before it is also described by an elliptic operator and it is completely analogous to the one solved numerically in section 5. with modified coefficients $D_i$ for $i=1,2,3$. It is possible to obtain an infinite set of discrete eigenvalues of the frequency $\omega_c$. 
\paragraph{Approximate spinning solutions.} A second case that we analyze corresponds to a spinning M2-brane with $f_c=r_c e^{i(k_c\sigma+l_c\rho)}$ with $r_c$ constant and $A_{r}$ is given by (\ref{sinusoidal}), in the approximation considered. The system is also satisfied and for this case the frequency in terms of the coefficients $D_i$  previously defined as
            \begin{eqnarray}
            {\omega_c^2 }
&=&
D_1\ k_c^2 
-D_2\ k_c l_c
+ D_3\ l_c^2  . 
            \end{eqnarray}
            
This ansatz has an associated well-defined symplectic gauge field with a nonvanishing curvature that contributes to the energy of the system.

In summary, under a first order approximation in both dynamical variables $Z_a$ and $A_{r}$ it is possible to obtain an approximate QBL solutions and  spinning solutions for the M2-brane with fluxes $C_{\pm}$ propagating on the three complex planes in the presence of  a well-defined nonvanishing small symplectic gauge field $dA$ field on its worldvolume with a curvature $\mathcal{F}\ne 0$.

\subsection{First order approximation on \texorpdfstring{$Z$}{}. }
Now we impose an approximation only to the $Z_a$ scalar fields. We will analyze an interesting case that corresponds to a membrane with different behaviour in the three complex planes for non-vanishing $A_{r}$.  We denote this case by 'mixed case'. In one complex plane, i.e. $Z_1$, we impose a  QBL ansatz where the approximation has been imposed. On the second complex plane $Z_2$ we assume a rotating ansatz and $Z_3$ and $X_2$ are assumed to be constant. On $Z_2$ and $A_{r}$ in distinction with the previous subsection, no approximation is imposed.
Before the approximation we assume the following embedding,
\begin{equation}\label{mixed case}
\begin{aligned}
Z_1&= f(\sigma, \rho) e^{i\Omega \tau}\ ,\\
Z_2&= r e^{i\beta}=r e^{i(k \sigma + l\rho + \omega \tau)}\ ,\\
A_{r}&= a_{r} \sigma + b_{r} \rho + c_{r} \tau \ ,
\end{aligned}    
\end{equation}
where $\Omega$, the frequency  for the 'breathing' mode of the membrane and $\omega$ the rotation frequency  in principle are assumed to be different.

The E.O.M reduce to 

\begin{eqnarray}
\label{1}
(Z_1):&\  \Omega^{2}f =&-E_{1}\left(\partial^2_{\sigma} f\right)+E_{2}\partial^2_{\sigma\rho} f-E_{3} \partial^2_{\rho} f,\\ 
\label{2}
(Z_2):&\ \omega^{2}=&
\sum_{r=9}^{10}\left[ k(R_{r}m_{r} + b_{r}) - l(a_{r}+R_{r}n_{r})\right]^2 + [ k(\partial_\rho f) - l (\partial_\sigma f) ]^2 \\
&& 
-i\left[ 
k (\partial_\rho f )(\partial^2_{\sigma\rho} f) 
-
l (\partial^2_\sigma f )(\partial_\rho f) 
-
k (\partial^2_\rho f) (\partial_\sigma f ) 
+
l(\partial^2_{\sigma\rho} f ) (\partial_{\sigma}f)
\right],\nonumber \\
\label{3}
(A_{r}):&\quad 0=& P_{r} \partial_\rho f \partial^2_\sigma f + Q_{r}\partial_\sigma f \partial^2_\rho f - 
\left[  Q_{r}\partial_\sigma f +P_{r}\partial_\rho f \right] \partial^2_{\sigma\rho} f \ ,
\end{eqnarray}

with
 \begin{eqnarray}
E_{1}&=&r^2l^2+\sum_{s=9}^{10}(m_{s} R_{s} + b_s)^2, \\
E_{2}&=&2\left(r^2kl+\sum_{s=9}^{10}(a_s + R_sn_{s})(b_s+R_sm_s), \right) \\
E_{3}&=&r^2k^2+\sum_{s=9}^{10}(n_{s} R_{s} + a_s)^2\ ,
\end{eqnarray}
and $P_{r}=m_{r}R_{r} + b_{r}$, $Q_{r}=n_{r}R_{r} + a_{r}$.
Under the above ansatz, the APD constraint {(\ref{constraint_f_g_general_for_Z})}  verifies directly, and central charge will restrict the wrapping numbers.
\paragraph{Approximation 'small' QBL}
 As before, we assume $Z_1= \lambda f_1(\sigma, \rho) e^{i\Omega \tau}$ and neglect the terms of order $O(\lambda^2)$. Since the $f_a$ functions modelling the QBL ansatz are coupled to the $A_{r}$, one can consider (\ref{3}) a restriction  on the equation (\ref{1}). The equation  (\ref{3}) for $A_{r}$  the linear ansatz trivially verifies without imposing any further condition.  The $Z_1$ equation (\ref{1}) for the approximate QBL ansatz has an analogous expression to the (\ref{qballaproximation}) with modified coefficients $E_i$ that now also depend on $A_{r}$ through the $a_{r}, b_{r}$ coefficients. The $Z_2$ equation, (\ref{2}) with the rotating ansatz can be explicitly solved for a constant modified rotation frequency determined by the values of the winding numbers, the moduli, the Fourier modes $k,l$ and the $A_{r}$ coefficients. 
 \begin{eqnarray}
 \Omega^{2}f =&-E_{1}\left(\partial^2_{\sigma} f\right)+E_{2}\partial^2_{\sigma\rho} f-E_{3} \partial^2_{\rho} f,\\ 
\omega^{2}=&
\sum_{r=9}^{10}\left[ k(R_{r}m_{r} + b_{r}) - l(a_{r}+R_{r}n_{r})\right]^2 \ .
\end{eqnarray}
See that due to the approximation performed  $Z_1$ and $Z_2$ EOM become disentangled. There is a rotating mode on the $Z_2$ complex plane  and a 'breathing' mode on the $Z_1$ plane. This solution also holds for  $A_{r}=constant$ particularizing the frequency to the values of $E_i$ with $a_{r}=b_{r}=0$. 
The $A_{r}$ field generically constrains the function $f$ allowed to model the QBL ansatz, but the constraint disappears when we assume the approximation. The validity of the linear $A_{r}$ ansatz in terms of an associated symplectic gauge field will be discussed at in section 7.3.

In summary, the mixed case considered here for the M2-brane with fluxes allows a behaviour of  approximate QBL on one plane and rotating in other for a  constant and linear $A_{r}$.

\paragraph{First order approximation on the \texorpdfstring{$A_{r}$}{}}
We have also explored other possibilities that involve to impose the approximation uniquely in the  gauge field $A_{r}$ for the case of a rotating ansatz on $Z_a$ like (\ref{2}) and a periodic regular function $A_{r}$ given by the equation (\ref{sinusoidal}), and we find that the system obliges to have zero central charge. Therefore, it does not represent an admissible solution for the M2-brane with fluxes.


\section{Analytical nontrivial \texorpdfstring{$A_{r}$}{} embeddings of the complete EOM system }
\label{S7}
In this section we will search for exact analytical solutions of the complete system of equations  admitting a nonvanishing $A_{r}$ for different ansätze of the $Z_a$. In particular we will analyze the cases where the complex scalar field $Z_a$ is a constant, a rotating solution or a QBL solution. In the subsection 7.3 we will discuss the validity of these solutions as  admissible components of the symplectic gauge field. 
\subsection{Case: Constant \texorpdfstring{$Z_a$}{} }
This configuration corresponds to a supermembrane with central  charges completely embedded in the compact sector that propagates as a point-like particle in the noncompact transverse space.  In this case the equations of motions and the APD constraint get extremely reduced to 
\begin{equation}
\label{Z constante}
\ddot{A}_{r}= \sum_{s=9}^{10}\left\{\mathcal{F}_{rs},X^{s}\right\}, \qquad\textrm{subject to} \quad \mathcal{D}_{r}\dot{A}_{r}=0 .
\end{equation}
\begin{itemize}
    \item{For embeddings associated to  a linear ansatz $A_{r}=a_{r}\sigma+b_{r}\rho+c_{r}\tau$ with $Z_a=constant$, we obtain that it satisfies the four set of equations and admits an infinite type of  solutions given by the real values of $a,b,c,d\in \mathbb{R}$.}
    \item{In the case of embeddings associated to a 'separable' ansatz $A_{r}(\sigma.\rho,\tau)=(a_{r}\sigma+b_{r}\rho)\tau+c_{r},$ the APD constraint imposes restrictions to the equations and in distinction with the ansatz previously analyzed, indeed it obliges $A_9=A_{10}$ with 
\begin{equation}
a= \left(\frac{R_9n_9+R_{10}n_{10}}{R_9m_9+R_{10}m_{10}}\right)b.    
\end{equation}
The system again allows to satisfy the nontrivial central charge condition since it constrains the real coefficient appearing in the $A_{r}$. In the analysis the stability of the solutions was guaranteed by imposing the equations to be preserved at each order of time $\tau$. This stability criteria restricts enormously the possibility of having polynomial ansätze of higher order in $\tau$.} 
    \item{One can analyze other solutions associated to the $A_{r}$ gauge field: In the case of  QBL ansatz or a spinning one on the $Z_a$, the first class APD constraint imposes a vanishing central charge.
Modifications of the preceding solution like for example  $A_9=r_9(\sigma, \rho)cos(\Omega\tau), A_{10}=r_{10}(\sigma, \rho)cos(\Omega\tau)$ with $r_9\ne r_{10}$ impose a zero frequency $\Omega$ and then constitute a subset of the first case analyzed for $c_{r}=0,$ and with
$A_9=A_{10}$. It trivially implies a vanishing central charge.}
\end{itemize}
\subsection{Case: Spinning membrane with nontrivial \texorpdfstring{$A_{r} $}{} field}
As we have already discussed in section 3.  there exists a subset of exact spinning M2 brane with flux solutions with constant $A_{r} $. Now we will generalize this result to include  solutions with nontrivial $A_{r} $ configurations.
Imposing a rotating ansatz on the complex scalar fields $Z_a $ with a nontrivial $A_{r} $

   \begin{equation}
\begin{aligned}
&Z_{a}(\tau, \sigma, \rho) = r_{a} e^{i \beta_{a}(\tau, \sigma, \rho)}, \quad \textrm{with}\quad  a=1, 2, 3;\quad 
\\
&X_{r}(\tau, \sigma, \rho) = R_{r}\left(n_{r} \sigma+m_{r} \rho\right)+q_{r} \tau + A_{r}(\tau, \sigma, \rho), \qquad \quad r=9, 10 \ .
\end{aligned}
\end{equation}
The precise equations of motion that one has to solve correspond to those shown in  the Appendix, once that $f_a(\sigma,\rho)=r_ae^{i(k_a\sigma_a+l_a\rho)}$.  As we have seen the system of PDE is very complex, and the constraints restrict enormously the possibilities to obtain nontrivial solutions.
However we find various interesting cases that solve the system. They correspond to 
\begin{itemize}
    \item{}The linear embedding case, $A_{r}=a_{r}\sigma+b_{r}\rho+c_{r}\tau$: These embeddings satisfy automatically all the set of equations fixing  the frequency of the rotating ansatz $\omega_c$ to the following value in terms of the coefficients $n_{r},m_{r}$, $k_a,l_a$

\begin{eqnarray}
\label{eom_for_Z_rotante_A_lineal_2}
\omega_{c}^{2}=\sum_{a=1}^{3} r_{a}^{2}  \left(k_{c} l_{a}-l_{c} k_{a}\right)^{2}
+
\sum_{s=9}^{10}\left[ R_s(m_sk_c-n_sl_c){+}(k_cb_s-l_ca_s) \right]^2 \ .
\end{eqnarray}

\item{} The 'separable' embedding case $A_{r}=(a_{r}\sigma+b_{r}\rho)\tau$ is more subtle. It requires to impose stability of the solutions such that the time dependence $\tau$ is cancelled order by order.
 This fact restricts enormously the equations allowed imposing also relation between the coefficients.
  \begin{equation}\label{chi}
     \chi\equiv\frac{a_9}{b_9}=\frac{a_{10}}{b_{10}}=\frac{k_c}{l_c}.
 \end{equation}
  The APD constraint again imposes restriction when we apply the condition (\ref{chi})
 \begin{equation}\label{RestricVinculoChi}
    \left(\chi m_{10}- n_{10}\right)=-\frac{R_{9}b_{9}}{R_{10}b_{10}}\left(\chi m_{9}- n_{9}\right),
\end{equation}
in such a way that there is a nontrivial relation between the coefficients of the complex variables $Z_a$ and those of the gauge field. Imposing (\ref{chi}) and (\ref{RestricVinculoChi}) the frequency acquires the following value,
\begin{equation}
\omega_{c}^{2}=l_{c}^{2} R_{9}^{2}\left(\chi m_{9}-n_{9}\right)^{2}\left[\left(\frac{b_{9}}{b_{10}}\right)^{2}+1\right] \ , 
\end{equation}               
satisfying the central charge condition. 
\item{}Other ansätze have been also explored: for example a rotating  or QBL ansätze on the $A_{r}$ are not allowed since they imply the vanishing of the central charge condition. 
\end{itemize}
\subsection{The symplectic gauge field}
\label{7.3}
So far we have analyzed several scenarios in the presence of different ansätze for  $A_{r}$: a constant, linear, separable, or regular periodic embedings. We have also commented that the QBL ansatz and rotating ansatz for the $A_{r}$ are not allowed since they imply a vanishing central charge. The symplectic gauge field is defined in terms of these functions $A_{r}(\sigma,\rho,\tau)$. Indeed it corresponds to $\mathbb{A}_{r}=\partial_a A_{r}d\sigma^a$. However in order to be well-defined, it must correspond to an exact one-form, whose symplectic curvature is topologically trivial. This property that characterize is relevant for the quantization process in the bosonic and the supersymmetric spectrum analysis.  The theory acquires a new single-valued dynamical degrees of freedom associated to a symplectic gauge field.

Clearly, the constant case and the regular periodic ansatz $A_{r}=a_{r}\sin{\sigma}+b_{r}\sin{\rho}+c_{r}\tau$ are single-valued with an exact associated one-form.  The regular periodic ansatz implies a well defined symplectic gauge field $\mathbb{A}=dA$ but it only satisfies the system in the approximations for the embeddings discussed in section 6. The constant ansatz of $A_{r}$ trivially implies a vanishing symplectic gauge field. 

However, the case of the linear and separable $A_{r}$ ansätze are different. They satisfy approximate equations of motion like occurs in the mixed case discussed in section 6.2.  They also satisfy exact analytical solutions to the fullfledged system of equations -where no approximation is imposed- for a constant and rotating $Z_a$ embedding.  Nevertheless, they have an associated multivalued one-form $\mathbb{A}$ and hence it cannot be interpreted as an admissible symplectic gauge field.  In other to correct it, the first thing to do is to impose periodicity on the $A_{r}$ function in order to be single-valued. Let us consider the linear case to illustrate it. 
We define the function as follows ,
 \begin{equation}\label{Solucionlineal}
A_{r}= a_{r}\sigma+b_{r}\rho+c_{r}\tau+d_{r}-2\pi a_{r}s_1(\sigma)-2\pi b_{r}s_2(\rho),\quad  \text{with} \quad \sigma,\rho \in \left[  2\pi i,2\pi (i+1)\right.] \quad i\in \mathbb{Z} \ .
 \end{equation}
 We have now a piecewise linear function that corresponds to a sawtooth wave where $s_1(\sigma)$ and $s_1(\sigma)$  represent the step functions. The function now is single-valued and periodic. 
However in order to be an admissible component of a symplectic gauge field, its associated one-form  $\mathbb{A}_{r}$   must be exact. That is, it must verify that 
\begin{equation}\label{exactness}
 \oint_{\mathcal{C}_s} \mathbb{A}_{r}= \oint_{\mathcal{C}_s}{dA_{r}}=0, \  \end{equation}
with $_{\mathcal{C}_s}$ the homological one-cycle basis of the target space 2-torus. Since the function  $A_{r}$ is single-valued but not regular, its derivative contains infinite delta functions, 
\begin{equation}
dA_{r}=\partial_a A_{r}d\sigma^a=a_{r}(1-2\pi \mathbf{m}(\sigma))d\sigma+b_{r}(1-2\pi \mathbf{m}{(\rho)})d\rho,
   \end{equation}
   being $\mathbf{m}(x)$ the shah function, also called Dirac comb,
   \begin{equation}
       \mathbf{m}(\sigma)=\sum_{s=-\infty}^{+\infty}\delta(\sigma-2s\pi),\quad  \mathbf{m}(\rho)=\sum_{q=-\infty}^{+\infty}\delta(\rho-2q\pi),\quad q, s\in \mathbb{N} \ .
   \end{equation}
   
Clearly, the exactness condition (\ref{exactness}) cannot be achieved for any value of $a_{r},b_{r}$. Consequently the symplectic curvature is also topologically nontrivial, something that is  excluded in the present theory.  Furthermore, the deltas would also appear in the E.O.M. with no clear interpretation of its meaning in terms of sources in the context of this theory.

In spite of these illnesses, one could try to define a regular associated field strength $\mathcal{F}$ associated to the linear $A_r$. Since its generic structure has the following form    
\begin{equation}
F_{9,10}=f_{9,10}+ B  \delta(\sigma-2\pi)-C\delta(2\rho-2\pi)+D\delta(\sigma-2\pi) \delta(2\rho-2\pi),  
\end{equation}
this could be achieved by imposing the coefficients $B=C=D=0$. For the case of the linear ansatz previously discussed, the coefficients can be explicitly computed
\begin{equation}
\begin{aligned}
&f_{9,10}=R_{9}(n_9b_{2}-m_{10}a_2)+R_{10}(n_{10}b_{1}-m_{10}a_1)+(a_2b_1-a_1b_2),\\
& B=2\pi[R_9m_9a_2+R_{10}m_{10}a_1-(a_2b_1-a_1b_2)],\\
& C=2\pi[R_9n_9b_2+R_{10}n_{10}b_1+(a_2b_1-a_1b_2)],\\
& D=4\pi^2(a_2b_1-a_1b_2)\ ,
 \end{aligned}
\end{equation}
but even in this case the vanishing of the non-regular part of the field strength automatically implies a vanishing central charge. Anyway, as we have already discussed, the linear  ${A}_{r}$ embedding does not represent an admissible degrees of freedom. 

\begin{figure}[ht]
    \centering
    \includegraphics[width=0.55\textwidth]{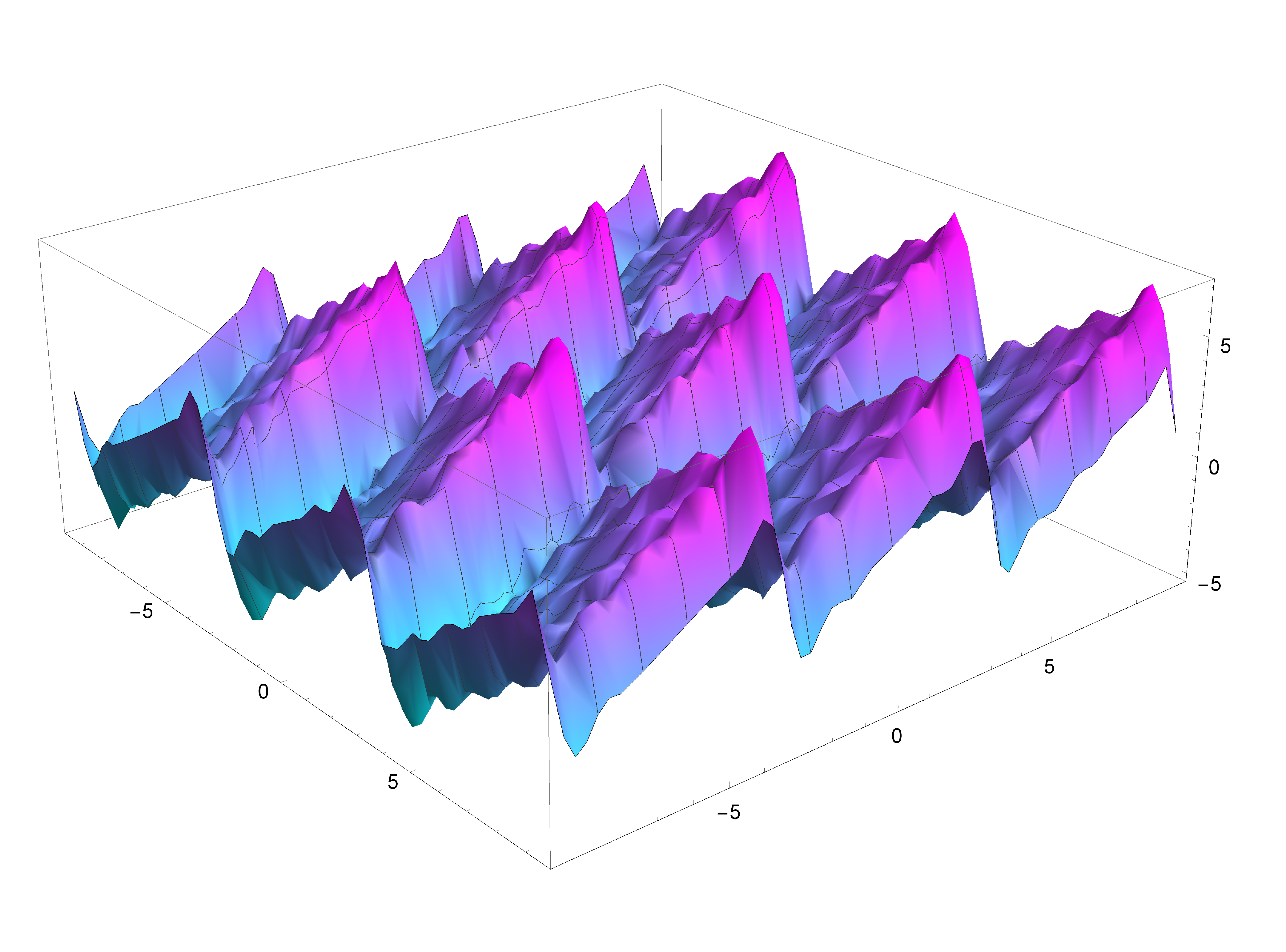}
    \caption{Expansion of the approximated function in one variable to order $N=4,\ M=4$ with $a_{r}=b_{r}=1$ and $d_{r}=0$, for a fixed time such that $c_{r}\tau=1$. Blue colors represent the lower values of the function. }
    \label{fig:ArLinealFouriern=4}
\end{figure}

The case of the separable ansatz on $A_{r}$ is even worse because of its time dependence that makes the associated one-form $dA_{r}$ to have a Dirac comb function for each time. Therefore although this solution mathematically satisfies the complete system of equations, it does not represent a physical solution. 
\paragraph{Fourier expansion} It is possible to obtain an approximate solution to the complete system of equations with a well defined associated symplectic gauge field. It corresponds to approximate the sawtooth function  representing the $A_r$ linear solution of (\ref{Solucionlineal}) by a truncated Fourier series in the multivalued functions in $(\sigma,\rho)$, see figure \ref{fig:ArLinealFouriern=4}. The coefficients of the expansion depend on $a_{r},b_{r}$ as follows
\begin{equation}
A_{r}(\sigma, \rho 
,\tau
) \sim  
\sum_{n=1}^{N}
2a_{r} \frac{(-1)^n}{n} \sin \left({ n }{} \sigma\right)
+
\sum_{m=1}^{M}
2b_{r} \frac{(-1)^m}{m} \sin \left({ m }{} \rho\right)
+
c_{r}
\tau +d_{r}\ ,
\end{equation}
for $N, M$ arbitrarily large finite numbers. 
For example see the Figure \ref{fig:ArLinealFouriern=4}. for the expanded function  to $N=4$ and $M=4$.  
In this case the function exhibits $n\times m$ local minima on each period. This solution automatically fulfills all of the conditions of periodicity, single-valued and an associated exact one-form, required to define properly the symplectic gauge field and consequently its associated curvature. The equations of motion are then approximately verified, with the approximation being better as the number $N$ of considered terms becomes higher but finite in the Fourier expansion.


\section{Discussion and Conclusions}
\label{s8}
We have obtained solutions to the classical equations of  the bosonic sector of the supermembrane theory formulated on  $M_9\times T^2$ in the presence of a quantized constant $C_{\pm}$ background. This non-vanishing constant three-form background induces a two-form flux in the target space that generates a worldvolume flux corresponding to a central charge condition. This condition has a topological nature, it is associated to the presence of monopoles on its worldvolume whose first Chern class characterizes the central charge integer. The so-called central charge condition is also equivalent to impose a nontrivial irreducible wrapping condition. It also exists a symplectic dynamical gauge field $\mathbb{A}_{r}$ defined on the worldvolume whose components are specified by the derivatives of the global functions $A_{r}$ discussed along the paper. The particularity of the associated supermembrane on this  background is that its supersymmetric description exhibit a purely discrete mass spectrum with finite multiplicity. The dynamics of well-defined quantum sectors of M-theory have not been previously studied.
Classically, the M2-brane with fluxes does not posses string-like configurations with zero energy cost and it is stable at quantum level. Hence the dynamics of the solutions describe an stable extended object. The solutions to the equations of motion of the M2-brane with fluxes are highly constrained because of the area preserving diffeomorphims and the flux condition. In spite of this, the mass operator of the bosonic M2-brane with fluxes for the particular spinning ansatz considered in section 4., when $A_{r}$ is assumed to be constant, contains the energy operator obtained in \cite{BRR2005nonperturbative}, hence the rotating membrane solutions found by the authors that also satisfy the central charge condition, are all M2-brane with fluxes solutions. We ave also obtained directly these solutions at the level of equations of motion of the M2-brane with $C_{\pm}$ fluxes. 

For $A_{r}$ constant we also explore 'Q-ball like' (QBL) ansatz  modelling the scalar complex variables. We obtain that in the absence of a worldvolume gauge field and approximating the QBL equations to first order, the  E.O.M describing the membrane admit an infinite set of solutions parametrized by a discrete frequency. We numerically  obtain the first nine eigenvalues and eigenvectors, for different boundary conditions. The first two boundary conditions corresponding to a periodic, and restricted ones are admissible for the toroidal membrane considered. The last one although it is an admissible solution of the PDE is not natural on the compact membrane. It is analyzed with illustrative purposes to compare the difference in the  breathing dependence of the membrane  with the boundary conditions.  

We also find that is possible to obtain a QBL solution and a spinning solution with non-vanishing gauge field when the analysis is performed by imposing approximations to second order in the parameter on the dynamical scalar fields $Z_a$ and $A_{r}$. Under these approximations we obtain solutions for a regular periodic trigonometric embedding $A_{r}$ that has a well-defined associated symplectic gauge field. We also consider less restrictive cases when the approximation is imposed in one of the two dynamical fields: We consider a mixed case in which all $Z_a$ behave differently, one is assumed constant, other rotating, and the third one that behaves with an approximated QBL ansatz, in the presence of a non-vanishing linear $A_{r}$. On this mixed case the system of equations is satisfied. The motion in the different planes decouples in the presence of the linear $A_{r}$. In the second case, the approximation is imposed only on $A_{r}$. We  consider a regular trigonometric embedding for the $A_{r}$ with $Z_a$ a rotating ansatz and the system only admits zero central charge and hence it does not represent an admissible solution to the M2-brane with fluxes.

In the cases in which no approximation is considered and the full-fledged equations are analyzed we obtain analytical solutions for the constant and rotating scalar complex field with simple polynomial embeddings $A_{r}$: linear or 'separable'. The APD and the topological constraints restrict enormously the embeddings allowed. An stability criteria on the solutions was imposed to guarantee that solutions are preserved at any time. This criteria is automatically fulfilled in the case of the linear ansatz on $A_{r}$,  but for the separable case it is only achieved when the coefficients of $A_{r}$ are restricted by a relation given by the wrapping numbers and the radii of the 2-torus. 

These linear  $A_{r}$ solutions are redefined in terms of a sawtooth function to guarantee the single-valuedness, but its derivative contains infinite number of deltas that renders the associated symplectic gauge field multivalued, an aspect that excludes them as suitable solutions. This solution to the complete system of E.O.M. can be approximated by their associated Fourier series truncated to a finite order. The approximation becomes increasingly better as more terms are included with $N$ restricted to be finite.  The 'separable' solution in spite of solving mathematically the system of equations does not admit an approximation to provide a sensible physical symplectic gauge field.

The results that we find do not exclude the possibility of obtaining exact and stable QBL membrane solutions with and without gauge field for the M2-brane with fluxes but its analysis out of the scope of the present paper. The existence of exact spinning solutions with a nontrivial symplectic gauge field seems strongly disfavoured as an exact analytical solution. It could be that  an underlying topological obstruction is behind this result, or that the assumption of rotation independence on the three complex planes is too restrictive, even the number and topology of the compact space dimensions may also play a role. In this paper we were interested in characterizing the spinning membrane solutions in the presence of flux $C_{\pm}$ on $M_9\times T^2$, other ansätze for the $Z_a$ complex scalar functions, different to  the ones analyzed in more general backgrounds, may admit exact symplectic gauge field configurations. 
These type of solutions deserve further study beyond the approximations. We plan to extend these results and analyze them in more interesting backgrounds.
\newline
One last comment is in order: It is well-known that accelerated point-like  charges always radiate even in the absence of any external field, but this is not necessarily the case for  accelerated  extended charges, \cite{Goedecke},\cite{Abbott}, or for stationary spinning solitons \cite{Volkov}. Hence, it is important to establish to what scenario our analysis corresponds. The solutions we find are stationary, in the sense that they do not generate any kind of bremsstrahlung or radiation effect. One type of solutions is spinning  and the other non-rotating Q-ball-like type. This nontrivial fact can be understood in the following way: As previously stated in section 2,  we characterize the dynamics of an 11D Supermembrane´s bosonic sector in a probe approximation described in three different but \textit{equivalent} ways: a) when it is irreducibly wrapped around the compact sector and propagates in a flat target space, $M_9\times T^2$. b) when it has a topological monopole charge  in the compact sector and propagates in a flat target space $M_9\times T^2$ and c) when the M2-brane propagates on a background  with a constant and quantized supergravity three-form on $M_9\times T^2$. All of them  share the same Hamiltonian and mass operator modulo a constant and the same equations of motion.

In the first scenario, it is clear that a single spinning  M2-brane that wraps irreducibly the compact sector, does not emit any kind of radiation nor presents any energy loss. Its center of mass, described by the zero modes, decouples from the nonzero ones and propagates as a free particle with a constant speed on $M_9\times T^2 $. Its momenta, angular momentum, and energy are conserved. Their spinning solutions -in the absence of gauge fields- were characterized by \cite{BRR2005nonperturbative}, and we reproduce the subset associated with the central charge condition. Since in this case, there is no energy lost, the same holds for the other descriptions, since all of them are equivalent. 

Lets try to explain better why this is the case:
In the case of a M2-brane with a monopole charge, this monopole is associated to the compact sector and furthermore the monopole does not rotate. The spinning solutions that we find are described by the complex embedding maps associated with the non-compact sector. The topological $U(1)$ monopole present in the M2-brane that we discuss has no dependence on time and it is characterized by its first Chern class. It is associated to a  $U(1)$ connection  $\widehat{A}$ constructed in terms of the harmonic pieces of the  the compact sector embedding maps which do not depend on time, only on the spatial worldvolume coordinates and  with $\widehat{A}_0=0$ by gauge fixing. Hence, there is no radiation associated with the monopole charge. The monopole condition we consider is completely equivalent to the central charge condition associated with the irreducibility of the wrapping of the M2-brane on the compact sector \cite{MARTINRestucciaTorrealba1998:StabilityM2Compactified}.

The third description is for spinning solutions associated with an M2-brane on a constant and quantized supergravity three-form $C_3$ on $M_9\times T^2$, which induces 2-form fluxes $C_{\pm}$. This background is the asymptotic limit of the supergravity background found by \cite{DUFF1Stelle991113} generated by an M2-brane source. The probe M2-brane that we consider is not charged under the $C_3$ potential since the charge associated with the background \cite{DUFF1Stelle991113} vanishes in the asymptotic limit and any probe must be consistent with the background. Hence, the picture we analyze corresponds to an uncharged spinning membrane propagating on a constant quantized three-form (vanishing 4-form flux) on $M_9\times T^2$  which do not generate any kind of backreaction. The theory is exactly equivalent to the previous two descriptions discussed, since it is connected to them through a canonical transformation modulo a constant shift.

 When a gauge field is present, which is the most general case, in any of the thee previous descriptions, we have found some approximate spinning solutions, see for example 6.1, and 6.2. The symplectic gauge field is described by $\mathbb{A}=dA$ with $A_r$, for $r=1,2$ the single-valued compact piece of the embedding map components. See that in this case, the symplectic gauge field has no time dependence either for the solutions found, since $A_r$  are linear in time. The remaining component in the covariant description $\mathbb{A}_0=\Lambda$ with $\Lambda$ the lagrange multiplier, is zero by gauge fixing as explained in section 2. Since $\mathbb{A}_r(\sigma,\rho)$ do not depend on time either, then the associated $\mathcal{F}_{0r}=0$  and  consequently, it cannot lead to any fluctuation in the symplectic gauge field. Since the symplectic  gauge field strength $\mathcal{F}$ is equal to a topological trivial U(1) gauge field $\mathcal{F}^{U(1)}=\mathcal{F}$, the same argument holds. Hence, there is not any kind of energy loss in agreement with the equivalence with the previous discussion in the absence of a dynamical gauge field. 
 
Interestingly, our analysis show certain resemblances with the results found in  \cite{Volkov} in the context of a non-abelian gauge field describing stationary spinning soliton solutions. These are obtained  when some conditions apply that also hold for the solutions  that we find: the presence of  several  complex scalars parametrizing the rotation with  their dependence on time entering through a phase, constant in time gauge fields, and axially symmetric solutions with a finite energy. For those cases they find several admissible solutions. An extension of this work in which we are interested is the search for 4D stationary soliton solutions of the M2-brane. It would be interesting to see if for more general cases, there exists any  connection  with the type of solutions found at effective energy level in \cite{Volkov}.

 \acknowledgments
 The authors are very grateful to A. Restuccia, P. Leon and C. Las Heras for helpful discussions or comments. We also thank to the referee for his/her comments that have help us to improve the paper. R.P. y J.M.P. thank to the projects ANT1956 y ANT1955 of the U. Antofagasta. P.D.A., M.P.G.M., J.M.P. and R.P. want to thank to SEM18-02 project of the U. Antofagasta. This work has been partially funded by Fondecyt grant 1180368. P.G. thanks to  BrainGain-Venezuela from Physics Without Frontiers program (ICTP), for kind support. M.P.G.M. and R.P. also thank to the international ICTP project NT08 for kind support.


\appendix
\section{Explicit EOM for the rotating ansatz}

The E.O.M for $A_{r}$ with $r,s=1,2$ and $r\ne s$ are,

\begin{eqnarray}
-\ddot A_{r} + (\sum_{a=1}^3k_a^2r_a^2 +n_{s}^2 R_{s}^2)\partial^2_\rho A_{r} +(-n_{r}n_{s}R_{r}R_{s}) \partial^2_\rho A_{s} +  \nonumber\\ 
2(-m_{s}n_{s}R_{s}^2-\sum_{a=1}^3 k_a l_a r_a^2) \partial_\rho\partial_\sigma A_{r} + 
R_{r}R_{s}(+n_{r}m_{s}+m_{r}n_{s})\partial_\sigma \partial_\rho A_{s}+ \nonumber\\
(\sum_{a=1}^3l_a^2r_a^2 +m_{s}^2 R_{s}^2) \partial_\sigma \partial_\sigma A_{r} - (m_{r}m_{s}R_{r}R_{s})  \partial_\sigma \partial_\sigma A_{s} + \nonumber \\
(-n_{s}R_{s}\partial_\sigma A_{r} - n_{r}R_{r}\partial_\sigma A_{s})\partial^2_\rho A_{s} +(2n_{s}R_{s})(\partial_\sigma A_{s})\partial^2_\rho A_{r} +\nonumber \\
(-2n_{s}R_{s}\partial_\rho A_{s}-2m_{s}R_{s}\partial_\sigma A_{s})\partial_\sigma\partial_\rho A_{r}+\nonumber \\
(n_{s}R_{s}\partial_\rho A_{r}+n_{r}R_{r}\partial_\rho A_{s}+m_{s}R_{s}\partial_\sigma A_{r}+m_{r}R_{r}\partial_\sigma A_{s})\partial_\rho\partial_\sigma A_{s}+\nonumber \\
2(m_{s}R_{s}\partial_\rho A_{s})\partial^2_\sigma A_{r}+
(-m_{s}R_{s}\partial_\rho A_{r}-m_{r}R_{r}\partial_\rho A_{s})\partial_\sigma \partial_\sigma A_{s}+\nonumber \\
-(\partial_\sigma A_{r})(\partial_\sigma A_{s})\partial^2_\rho A_{s}+(\partial_\sigma A_{s})^2\partial^2_\rho A_{r}-2(\partial_\rho A_{s})(\partial_\sigma A_{s})\partial_\sigma\partial_\rho A_{r}+\nonumber \\
(\partial_\rho A_{s}\partial_\sigma A_{r}+\partial_\rho A_{r}\partial_\sigma A_{s})\partial_\sigma\partial_\rho A_{s}+(\partial_\rho A_{s})^2\partial^2_\sigma A_{r}-(\partial_\rho A_{r}\partial_\rho A_{s})\partial^2_\sigma A_{s}=0\ . \nonumber
\end{eqnarray}
For $Z_c$
\begin{eqnarray}
\label{eom_for_Z1_general}
-\omega_{c}^{2}+
\sum_{a=1}^{3} r_{a}^{2}  \left(k_{a} l_{c}-l_{a} k_{c}\right)^{2}+
R_{9}^{2} \left(n_{9} l_{c}-m_{9} k_{c}\right)^{2}
+
R_{10}^{2}\left(n_{10} l_{c}-m_{10} k_{c}\right)^{2}+ \nonumber \\
k_c^2[(\partial_\rho A_9)^2 +(\partial_\rho A_{10})^2 ]+ l_c^2[(\partial_\sigma A_9)^2 +(\partial_\sigma A_{10})^2 ]+\nonumber \\
-2k_cl_c[(\partial_\rho A_9)(\partial_\sigma A_9)+(\partial_\rho A_{10})(\partial_\sigma A_{10})]+\nonumber \\
i[k_c(\partial_\sigma A_9\partial^2_\rho A_9 + \partial_\sigma A_{10}\partial^2_\rho A_{10}-\partial_\rho A_{9}\partial_\sigma\partial_\rho A_{9}-\partial_\rho A_{10}\partial_\sigma\partial_\rho A_{10})+ \nonumber\\ 
+l_1(-\partial_\sigma A_9\partial_\rho\partial_\sigma A_9 - \partial_\sigma A_{10}\partial_\sigma\partial_\rho A_{10}+\partial_\rho A_{9}\partial^2_\sigma A_{9}+\partial_\rho A_{10}\partial^2_\sigma A_{10})]+ \nonumber\\
2k_cR_9(k_cm_9-n_9l_c)\partial_\rho A_{9}+2k_cR_{10}(k_cm_{10}-n_{10}l_c)\partial_\rho A_{10}+\nonumber\\ +(-2k_cm_9R_9l_c+2n_9R_9l_c^2)\partial_\sigma A_{9}+(-2k_cm_{10}R_{10}l_c+2n_{10}R_{10}l_c^2)\partial_\sigma A_{{10}}+\nonumber\\ i[k_cn_9R_9\partial_\rho \partial_\rho A_{9}+k_cn_{10}R_{10}\partial_\rho \partial_\rho A_{10}+(-k_cm_9R_9-l_cn_9R_9)\partial_\sigma \partial_\rho A_{9}+\nonumber\\ (-k_cm_{10}R_{10}-l_cn_{10}R_{10})\partial_\sigma \partial_\rho A_{10}+l_cm_9R_9\partial_\sigma \partial_\sigma A_{9}+l_cm_{10}R_{10}\partial_\sigma \partial_\sigma A_{10}] =0 \ .\nonumber
\end{eqnarray}

The APD constraint once the rotating ansatz has been substituted becomes reduced to  
\begin{eqnarray}
(\partial_\sigma \dot{A}_{r} \partial_\rho \hat{X}^{r} -  \partial_\rho \dot{A}_{r} \partial_\sigma \hat{X}^{r})+(\partial_\sigma \dot{A}_{r} \partial_\rho A^{r} -  \partial_\rho \dot{A}_{r} \partial_\sigma A^{r})=0 .
\end{eqnarray}

\bibliography{referencesURL}

\end{document}